\documentclass[twoside,twocolumn,english,prb]{revtex4}
\usepackage[T1]{fontenc}
\usepackage[latin9]{inputenc}
\usepackage{amssymb}
\usepackage{graphicx}
\usepackage{wasysym}
\usepackage{esint}

\makeatletter
\@ifundefined{textcolor}{}
{%
 \definecolor{BLACK}{gray}{0}
 \definecolor{WHITE}{gray}{1}
 \definecolor{RED}{rgb}{1,0,0}
 \definecolor{GREEN}{rgb}{0,1,0}
 \definecolor{BLUE}{rgb}{0,0,1}
 \definecolor{CYAN}{cmyk}{1,0,0,0}
 \definecolor{MAGENTA}{cmyk}{0,1,0,0}
 \definecolor{YELLOW}{cmyk}{0,0,1,0}
}

\makeatother

\usepackage{babel}
\begin{document}
\title{Spin bath dynamics and dynamical Renormalization Group}
\author{Álvaro Gómez-León}
\affiliation{Instituto de Ciencia de Materiales de Madrid (ICMM-CSIC), Cantoblanco,
28049 Madrid, Spain}
\date{\today}
\begin{abstract}
We discuss the quantum dynamics of the central spin model in a regime
where the central spin and bath are slaved to each other. The exact
solution is found when the bath is static, and is compared with the
effect of an external field, finding that they are inequivalent due
to the quantum nature of the environment. When the bath has dynamics,
we analyze the differences between the numerical simulation using
time-dependent perturbation theory and the equation of motion technique,
which shows better accuracy. We demonstrate that the use of dynamical
Renormalization Group (dRG), simultaneously with the equation of motion
technique, provides a suitable analytical tool to understand the physics,
to capture the main physical processes, and a powerful method to eliminate
secular terms. In addition, this approach allows to separate classical
non-linear behavior from corrections due to quantum correlations.
\end{abstract}
\maketitle

\section{Introduction}

During the last decade, a growing interest in the understanding of
dynamical quantum systems has emerged. Motivated from both, theory
and experiment, a whole new area of physics is being developed, where
quantum systems and their dynamics play a dominant role. While non-interacting
systems are quite well understood, interacting systems can display
exotic new physics such as Floquet phases\citep{FloquetTI,FFCI,ObservationFTI},
time crystals\citep{Time-Crystal-Wilczek,Time-Crystals-Review}, many-body
localization\citep{Many-Body-Loc1,Many-Body-Loc2} and complex dynamics\citep{SpinBathSimulations,UnconventionalRates}.
While the simulation of classical complex systems with non-linearities
can be challenging (weather forecast, stock-market predictions, social
behavior or swarming), an extra difficulty arises in quantum systems,
due to the presence of entanglement.

The central spin model is one of the canonical models to study the
dynamics of quantum interacting systems\citep{TheoryBathSpin2000}.
It describes a quantum spin interacting with a set of localized modes,
and can mimic molecular magnets interacting with impurities\citep{MolecularMagnetsDecoherence},
flux qubits coupled to electric dipoles\citep{MartinisSQUID} and
many other effective two-level systems interacting with localized
modes. Interestingly, the dynamics in this model can be quite complex,
as it is known that the localized nature of these modes requires a
non-perturbative analysis and gives rise to a rich dynamical behavior\citep{Crossover},
which can be quite different from the spin-boson model\citep{SpinBosonModel}.
For example, in some regimes the bath dynamics is slaved to the motion
of the central spin, and the memory of the bath becomes important.
This is typically discussed in terms of the dimensionless parameter
$\left|A_{i}\right|/\left|\vec{\Delta}\right|$, where $A_{i}$ denotes
the coupling strength between the central system and the i-th bath
spin, and $\vec{\Delta}$ is the external field that gives free dynamics
to the bath spins\citep{TheoryBathSpin2000}. Therefore, if the coupling
between the two systems dominates over the free Hamiltonian for the
bath $\left|A_{i}\right|/\left|\vec{\Delta}\right|\gg1$, the dynamics
of the two systems is highly correlated.

In this work we study the dynamics of the central spin model in the
regime where the dynamics of the central spin and bath are slaved
to each other. First, we exactly solve the case of a static bath and
demonstrate that the spin bath is not equivalent to an external magnetic
field, specially when the bath is not in its ground state. This leads
to a damping of coherent oscillations at short times, which can be
confused with decoherence (however, in this case entanglement between
the two systems is not formed and it is purely a dephasing effect,
which can be reversed using spin echo). Then we discuss the regime
where the bath is dynamical, and the new mechanisms which can modify
the coherent oscillations at different time-scales. In particular,
we show how non-perturbative contributions from many-body effects
lead to instanton-like transitions in the central spin, and to a suppression
of the coherent oscillations due to the formation of correlations
with the bath. These mechanisms appear at very different time-scales
and can be captured numerically and analytically. This is possible
due to the use of dynamical Renormalization Group (dRG), which allows
for a natural time-scale separation, when combined with the equation
of motion technique.

\section{Model}

We consider the next Hamiltonian describing a two-level system (or
qubit) interacting with a bath of spins $\vec{I}_{i}$ of arbitrary
spin value $P_{i}$:
\begin{eqnarray}
H & = & H_{0}+V_{B}\label{eq:H}
\end{eqnarray}
being
\begin{eqnarray}
H_{0} & = & -B_{z}S^{z}-B_{\perp}S^{x}-\sum_{i=1}^{N}\left(\Delta_{z}-S^{z}A_{i}\right)I_{i}^{z}\\
V_{B} & = & -\sum_{i=1}^{N}\Delta_{\perp}I_{i}^{x}
\end{eqnarray}
We have assumed that the interaction is purely longitudinal (typically
due to a large crystal field anisotropy), and that the central spin
and bath spins couple, in addition to the longitudinal fields $B_{z}$
and $\Delta_{z}$, to the transverse fields $B_{\perp}$ and $\Delta_{\perp}$,
respectively.

If the interaction $A_{i}$ dominates, the central spin couples to
a longitudinal Overhauser field produced by the bath (it can be experimentally
quite large, as $A_{i}$ does not scale as $N^{-1/2}$, which would
be the case for delocalized modes), while each bath mode couples to
a weaker field, produced by the central spin only. When the transverse
fields are added, bath and central spin precess at different rates,
and spin-flip transitions can happen, mediated by the interaction.
Different Hamiltonians with more general couplings can also be studied
using this formalism, but Eq.\ref{eq:H} has the necessary ingredients
to produce interesting effects and simple analytical expressions.

\section{Exact solution for a static bath}

For $\Delta_{\perp}=0$ the Hamiltonian reduces to $H_{0}$ in Eq.\ref{eq:H},
and in this case, the model can be exactly solved and displays interesting
features. In order to find the solution, it is useful to consider
the many-body basis $|M;\vec{P},\vec{m}\rangle$, with $M=\pm1/2$
labeling the two states of the central spin, and where $\vec{P}=\left(P_{1},\ldots,P_{N}\right)$
and $\vec{m}=\left(m_{1},\ldots,m_{N}\right)$ are $N$-dimensional
vectors labeling the values of the bath spins at the different sites
and their projection onto the z-axis, respectively. The calculation
of the magnetization is straightforward because the system becomes
block-diagonal for different spin bath configurations $\vec{m}$.
For example, the time evolution of the longitudinal magnetization,
assuming that initially the central spin is in an eigenstate of $S^{z}$,
yields (full expression and derivation in the Appendix):
\begin{equation}
S^{z}\left(t\right)=\sum_{\vec{P},\vec{m}}S_{\vec{m}}^{z}\left(1+B_{\perp}^{2}\frac{\cos\left(\Omega_{\vec{m}}t\right)-1}{\Omega_{\vec{m}}^{2}}\right)\label{eq:Exact-Dynamics1}
\end{equation}
where we have defined the central spin frequency for a given bath
configuration $\vec{m}$ as: 
\begin{equation}
\Omega_{\vec{m}}=\sqrt{B_{\perp}^{2}+\left(B_{z}-\vec{A}\cdot\vec{m}\right)^{2}}
\end{equation}
and $S_{\vec{m}}^{z}=\sum_{M}M|M;\vec{P},\vec{m}\rangle\langle M;\vec{P},\vec{m}|$
is the $S^{z}$ operator for the bath configuration $\vec{m}$.

Eq.\ref{eq:Exact-Dynamics1} has very interesting features, some of
which have been discussed in \citep{Zurek-SpinBath2005}. In this
case, the expression applies for arbitrary bath spin values $P_{i}$
and to any initial state configuration (e.g., this expression can
be applied if the spin bath contains different nuclear isotopes).
The main feature is the summation over all spin bath configurations
$\sum_{\vec{P},\vec{m}}$ which can radically modify the central spin
dynamics, depending on the initial condition for the bath. In this
case, the initial state preparation becomes quite relevant for the
subsequent dynamics.

When the bath is at low temperature $T\ll\left|A_{i}\right|$, $\left|\vec{\Delta}\right|$,
mostly the ground state will be occupied and the summation over bath
configurations reduces to a single term which has a shifted Zeeman
splitting $B_{z}\rightarrow B_{z}-\vec{A}\cdot\vec{m}$. On the other
hand, for many experimental setups the interaction with each spin
is weak, and although the central spin will be at low temperature,
the bath will be in the high temperature regime $\left|\vec{B}\right|\gg T\gg\left|A_{i}\right|$,
$\left|\vec{\Delta}\right|$. This implies that almost all hyperfine
levels will be equally occupied, and the sum over bath configurations
has many terms, where each of them contributes with a different frequency
$\Omega_{\vec{m}}$. In this case, although the system has a Poincare
recurrence time at long times $\tau_{rec}\sim\left|A_{i}\right|^{-1}$,
the dynamics resembles a ``decoherence'' process due to the bath.
In many cases, the hyperfine levels are close to each other and broadening
is large enough as to make them overlap. Then, one can approximate
the sum over bath configurations by an integral with a density of
states $J\left(\alpha\right)=\sum_{\vec{m},\vec{P}}\delta\left(\alpha-B_{z}+\vec{A}\cdot\vec{m}\right)$,
and calculate its contribution using a stationary phase approximation.
This transforms Eq.\ref{eq:Exact-Dynamics1} into:
\begin{equation}
S^{z}\left(t\right)\simeq S^{z}\left(1+B_{\perp}^{2}\int_{-\infty}^{\infty}J\left(\alpha\right)\frac{\cos\left(\Omega_{\alpha}t\right)-1}{\Omega_{\alpha}^{2}}d\alpha\right)
\end{equation}
where we have defined $\Omega_{\alpha}=\sqrt{B_{\perp}^{2}+\alpha^{2}}$,
and the spectral function is given by:
\begin{eqnarray}
J\left(\alpha\right) & = & \frac{e^{-\frac{\left(\alpha-B_{z}\right)^{2}}{2\sigma^{2}}}}{\sqrt{2\pi\sigma^{2}}},\ \sigma=\sqrt{\frac{1}{6}\sum_{i}A_{i}^{2}P_{i}^{2}\frac{P_{i}+4}{P_{i}+2}}
\end{eqnarray}
for the case of large $P_{i}$ bath spins. The assumption of large
$P_{i}$ is not required, but simplifies the expressions (the general
case is analyzed in the Appendix).

\begin{figure}
\includegraphics[scale=0.48]{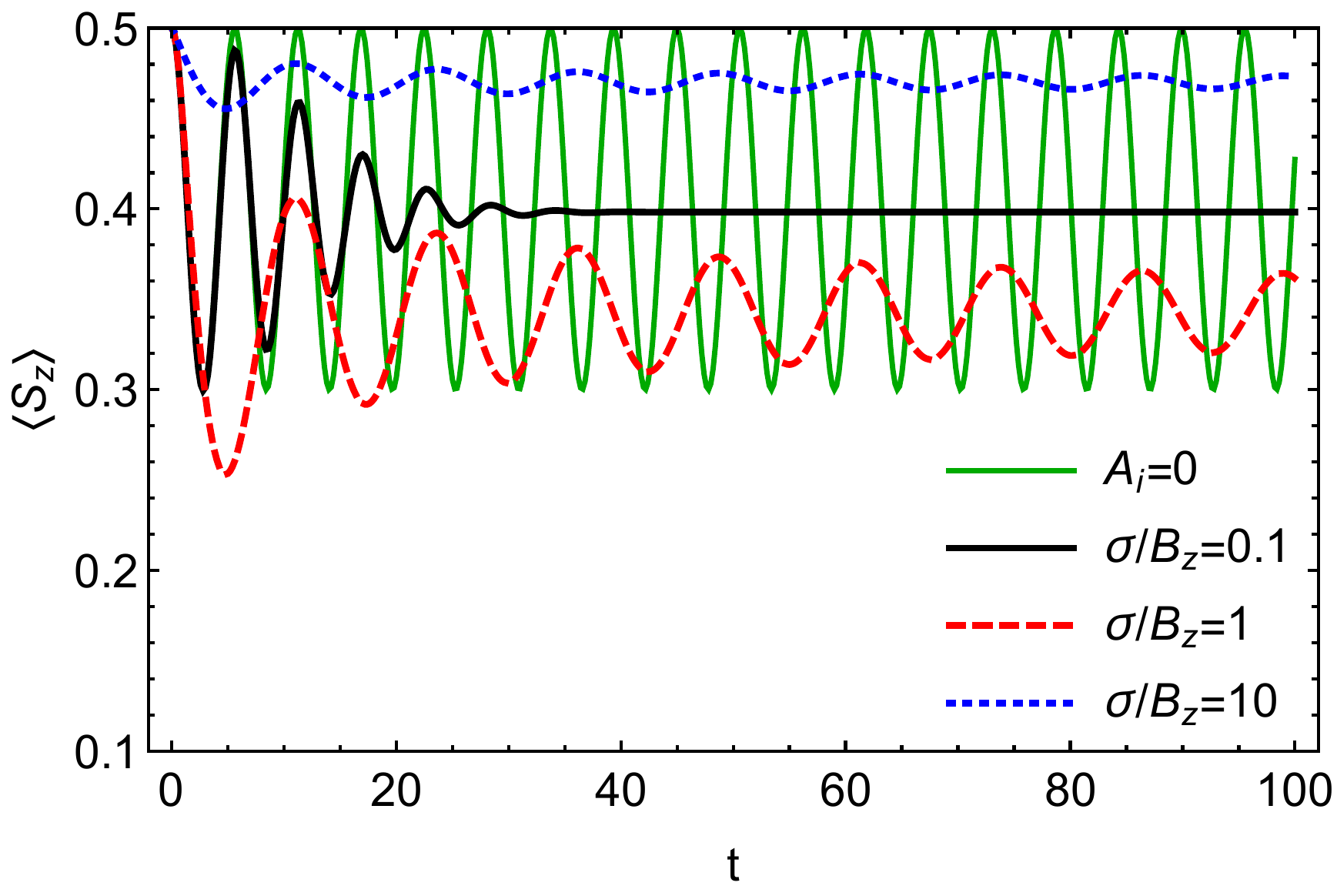}

\caption{\label{fig:Comparison0}Comparison between the free dynamics and the
case with a bath of spins for different spectral functions $J\left(\alpha\right)$
characterized by $\sigma$. We have chosen $B_{\perp}/B_{z}=0.5$
and an initial spin up state.}
\end{figure}

Then, the dynamics is governed by a Gaussian distribution peaked at
$B_{z}$, which broadens with the number of bath spins as $N^{1/2}$,
and linearly with $A_{i}$ and $P_{i}$. Importantly, this happens
even for the case of ordered couplings $A_{i}=A$, indicating that
it is purely a bath effect, and not a disorder average. Fig.\ref{fig:Comparison0}
shows a comparison between the free dynamics and the case with a spin
bath for different values of $\sigma$.
It shows that the coherent oscillations produced by the transverse
field are damped due to a phase interference between the different
bath configurations. Furthermore, the bath controls both, the damping
of the oscillations and the average value of the longitudinal magnetization.

It is important to realize that the reason why the coherent oscillations
are suppressed is because the bath does not act as a classical magnetic
field, and its quantum nature allows different bath configurations
to evolve with different phases, resulting in the suppression of coherent
oscillations. Interestingly, when the width of the bath distribution
$\sigma$ is of the order of the splitting $B_{z}$, the coherent
oscillations remain for a long time. This shows that by tuning $B_{z}$
one can minimize the effect of the bath, or by studying the time evolution
as a function of $B_{z}$, extract information about the density of
states (DOS) of the environment.

\section{Dynamical bath}

\begin{figure}
\includegraphics[scale=0.48]{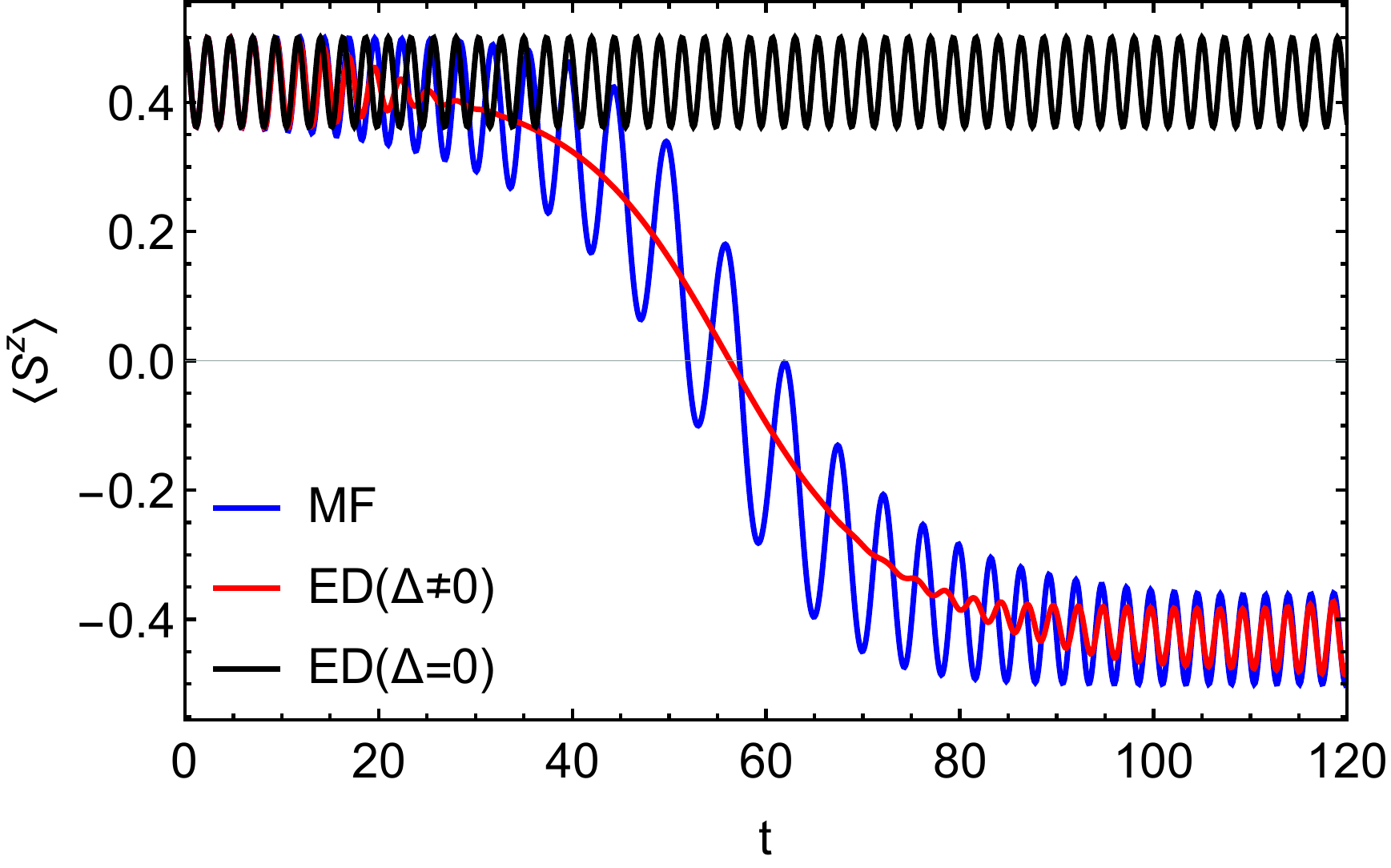}

\caption{\label{fig:Numerical1}Comparison between the exact dynamics for $\Delta_{\perp}/B_{\perp}=0$
(black) and $\Delta_{\perp}/B_{\perp}=0.03$ (red), for the case of
homogeneous couplings $A/B_{\perp}=0.05$, $N=100$ and $B_{z}=\Delta_{z}=0$.
The dynamics of the bath spins, produced by $\Delta_{\perp}\protect\neq0$,
leads to instanton-like transitions in the central spin at long time-scales,
when $N\gg1$ (this transition gets more abrupt as $N$ increases).
(Blue) Numerical solution of the mean field equations. When $\Delta_{\perp}/B_{\perp}\ll1$
the short time dynamics is identical to the case $\Delta_{\perp}=0$,
with a static bath. The initial condition is a product state with
central spin up and all bath spins up.}
\end{figure}

When the bath couples to a transverse field ($\Delta_{\perp}\neq0$),
the longitudinal magnetization for the bath becomes time-dependent
as well. This makes both systems precess at different rates, as typically
the central spin dynamics is much faster. However, as their time evolution
is not independent, a resonance can happen at longer time-scales,
and produce substantial changes in the dynamics of the central spin.
This is shown in Fig.\ref{fig:Numerical1}, where we have calculated
the exact time-evolution for the longitudinal magnetization of the
central spin, for an initially fully polarized bath (with this initial
condition the previously discussed ``false decoherence'', induced
by the sum over bath configurations, is absent. Therefore, all changes
are a consequence of the bath dynamics). The black line shows the
static bath case ($\Delta_{\perp}=0$), and the red line shows the
case with a small transverse field ($\Delta_{\perp}/B_{\perp}\ll1$).
When the bath is static, the central spin coherently oscillates with
amplitude proportional to $B_{\perp}/\left(B_{z}-\vec{A}\cdot\vec{m}\right)$
and frequency $\Omega_{\vec{m}}$; however, when the bath is dynamical
two main effects can happen at different time scales:
\begin{enumerate}
\item The amplitude of the oscillations gets damped at short time-scales.
\item The central spin magnetization flips at long time-scales.
\end{enumerate}
The first effect is a consequence of the formation of entanglement
between the central spin and the bath spins, while the second effect
is produced due to a resonance between the central spin and the bath.

\subsection{Perturbation theory}

As a first approach, let us consider time-dependent perturbation theory
around the unperturbed solution (i.e., for a static bath with $\Delta_{\perp}=0$).
A general state $|\Psi\left(t\right)\rangle$ can be expressed in
this basis as:
\begin{equation}
|\Psi\left(t\right)\rangle=\sum_{M,\vec{P},\vec{m}}c_{M,\vec{P},\vec{m}}\left(t\right)e^{-itE_{\vec{m}}^{M}}|M,\vec{P},\vec{m}\rangle
\end{equation}
where $H_{0}|M,\vec{P},\vec{m}\rangle=E_{\vec{m}}^{M}|M,\vec{P},\vec{m}\rangle$
and
\begin{equation}
E_{\vec{m}}^{M}=-\Delta_{z}\sum_{i}m_{i}+M\Omega_{\vec{m}}
\end{equation}
From the time-dependent Schrödinger equation one finds that the time
evolution is given by\footnote{In what follows we do not write explicitly the dependence on $\vec{P}$,
however it must be considered when calculating observables.}:
\begin{equation}
\dot{c}_{M,\vec{m}}\left(t\right)=-i\sum_{\vec{m}^{\prime}}c_{M,\vec{m}^{\prime}}\left(t\right)e^{-it\left(E_{\vec{m}^{\prime}}^{M}-E_{\vec{m}}^{M}\right)}\langle\vec{m}|V_{B}|\vec{m}^{\prime}\rangle\label{eq:Perturbative1}
\end{equation}
where we have used that $V_{B}$ does not change the central spin
state $M$ or the bath spin $\vec{P}$. The matrix elements can be
calculated straightforwardly:
\begin{equation}
\langle\vec{m}|V_{B}|\vec{m}^{\prime}\rangle=-\frac{\Delta_{\perp}}{2}\sum_{i=1}^{N}\left(\gamma_{P_{i},m_{i}^{\prime}}\delta_{\vec{m}-1_{i},\vec{m}^{\prime}}+\gamma_{P_{i},m_{i}}\delta_{\vec{m}+1_{i},\vec{m}^{\prime}}\right)
\end{equation}
where
\begin{equation}
\gamma_{P_{i},m_{i}}=\sqrt{P_{i}\left(P_{i}+1\right)-m_{i}\left(m_{i}+1\right)}
\end{equation}
and $\vec{m}\pm1_{i}$ in $\delta_{\vec{m}\pm1_{i},\vec{m}^{\prime}}$
corresponds to the bath configuration $\vec{m}$ with the bath spin
at the $i$-th site changed by a unit.

For $N\gg1$ the system of equations cannot be exactly solved, but
one can use a perturbative expansion:
\begin{equation}
c_{M,\vec{m}}\left(t\right)=c_{M,\vec{m}}^{\left(0\right)}+\Delta_{\perp}c_{M,\vec{m}}^{\left(1\right)}\left(t\right)+\Delta_{\perp}^{2}c_{M,\vec{m}}^{\left(2\right)}\left(t\right)+\ldots
\end{equation}
and solve Eq.\ref{eq:Perturbative1} for different orders of $\Delta_{\perp}$.
We have calculated the solution up to second order in $\Delta_{\perp}$
to try to reproduce the results from Fig.\ref{fig:Numerical1}. To
first order in $\Delta_{\perp}$, the solution couples the states
$|M,P\rangle$ and $|M,P-1\rangle$. To second order in $\Delta_{\perp}$,
the state $|M,P-2\rangle$ also weakly couples to $|M,P\rangle$,
but one also finds the next secular term in the solution:
\begin{equation}
ic_{M,\vec{m}}^{\left(0\right)}\left(t_{0}\right)\left(t-t_{0}\right)\left(\frac{\Delta_{\perp}}{2}\right)^{2}f\left(\vec{m}\right)\label{eq:secular-term}
\end{equation}
where
\begin{equation}
f\left(\vec{m}\right)=\sum_{i=1}^{N}\left(\frac{\gamma_{P_{i},m_{i}-1}^{2}}{E_{\vec{m}}^{M}-E_{\vec{m}-1_{i}}^{M}}+\frac{\gamma_{P_{i},m_{i}}^{2}}{E_{\vec{m}}^{M}-E_{\vec{m}+1_{i}}^{M}}\right)
\end{equation}
At this point it is interesting to introduce the technique of dynamical
Renormalization Group and the physical reason behind the appearance
of secular terms: Secular terms are common in perturbative expansions.
Mathematically, they produce a cut-off beyond which the perturbative
solution is not valid, and in the present case this happens for times
$t-t_{0}\apprge\Delta_{\perp}^{-2}$. They are produced by resonant
terms in the perturbative solutions, and it can be seen that physically,
the same principle applies: \textit{Resonant physical processes lead
to secular terms in the perturbative expansion}\footnote{Notice that here the resonance comes from the two step process of
flipping back and forth a bath spin, which leaves the energy unchanged.
This can be done for all bath spins, even in the disordered case,
which is why it can be an important correction}. The main reason is that resonant terms produce large corrections
which are non-perturbative, and when tried to be expressed in a perturbative
way, they restrict the validity of the expansion. Hence, secular terms
give important information about large corrections to perturbative
solutions, and lead to the emergence of new time-scales. Therefore,
it is important to find a way to deal with them and to extract this
information. This is what dRG does, by encoding the secular terms
in the boundary conditions.

In order to understand the basic idea, let us consider the previous
second order solution with a secular term (Eq.\ref{eq:secular-term}).
The secular term dominates when $t-t_{0}\gtrsim\Delta_{\perp}^{-2}$,
and it would be interesting if one could keep $t-t_{0}$ always small.
This can be done by assuming that $t_{0}$ is dynamical, but the price
to be paid is that the boundary conditions also become dynamical (because
they are functions of $t_{0}$). As the total solution cannot depend
on this arbitrary cut-off, one must impose the condition:
\begin{equation}
\partial_{\tau}c_{M,\vec{m}}\left(t\right)=0
\end{equation}
Where we have substituted $t_{0}\rightarrow\tau$, to indicate that
$t_{0}$ is now a dynamical variable. This produces a flow equation
for the boundary condition, and by choosing $\tau=t$, one can eliminate
the secular term, which is now encoded in the time dependence of the
boundary condition\citep{dRG-Physics,dRG-Sarkar2011}. This approach
gives similar results to Multiple-scale analysis, also well known
in the literature. In that case one just needs to impose an ansatz
with different time-scales for the solutions $c\left(t\right)\rightarrow c\left(t,\tau_{1},\tau_{2},\ldots\right)$,
being $\tau_{n}=\Delta_{\perp}^{n}t$\citep{dRG-QuantumHarmonicOscillator1996,dRG-QuantumOptics2003}.

When this formalism is applied to the present problem, one finds that
up to second order in $\Delta_{\perp}$, the boundary condition changes
as:
\begin{equation}
\partial_{\tau}c_{M,\vec{m}}^{\left(0\right)}\left(\tau\right)\simeq ic_{M,\vec{m}}^{\left(0\right)}\left(\tau\right)\left(\frac{\Delta_{\perp}}{2}\right)^{2}f\left(\vec{m}\right)
\end{equation}
Its solution is a shift in the frequency of the coherent oscillations,
proportional to $f\left(\vec{m}\right)$, and a small damping of oscillations
due to the interference between different states, but the solution
does not capture the instanton-like transition. The reason is that
the instanton-like transition involves the inversion of the all the
bath spins, and therefore must include states with $\vec{m}=-\vec{P}$.
We show next that one can capture it starting from a non-linear set
of equations of motion.

\subsection{Mean field solution}

In order to capture the instanton-like solution, we go to the Heisenberg
picture and calculate the dynamics using the equation of motion $\partial_{t}\hat{O}=i\left[H,\hat{O}\right]$
for the spin operators, being $\hat{O}$ an arbitrary operator. For
the central spin one finds ($\epsilon_{\alpha\beta\delta}$ is the
Levi-Civita symbol and greek indices correspond to the three spatial
axis):
\begin{equation}
\partial_{t}S^{\alpha}=\sum_{\mu,\theta=x,y,z}\epsilon_{\mu\alpha\theta}\left(B_{\mu}-\sum_{i=1}^{N}A_{i}^{\mu}I_{i}^{\mu}\right)S^{\theta}
\end{equation}
while for the bath spins one finds:
\begin{equation}
\partial_{t}I_{i}^{\alpha}=\sum_{\mu,\theta=x,y,z}\epsilon_{\mu\alpha\theta}\left(\Delta_{\mu}-S^{\mu}A_{i}^{\mu}\right)I_{i}^{\theta}\label{eq:BathEOM}
\end{equation}
This set of coupled equations are general for a wide number of Hamiltonians,
but in this work we focus on the specific case of Eq.\ref{eq:H},
with $\vec{B}=\left(B_{\perp},0,B_{z}\right)$, $\vec{\Delta}=\left(\Delta_{\perp},0,\Delta_{z}\right)$
and $\vec{A}_{i}=\left(0,0,A_{i}\right)$. In order to illustrate
the emergence of new time-scales, we first consider a mean field decoupling
of the equations. This implies that correlations between spins are
neglected, and the product of spin operators is substituted by the
product of their individual average value $\langle I_{i}^{\alpha}S^{\beta}\rangle\simeq\langle I_{i}^{\alpha}\rangle\langle S^{\beta}\rangle$
with respect to an initial density matrix $\rho_{0}$ describing the
initial state of the system. The numerical solution is shown in Fig.\ref{fig:Numerical1}
(blue), and it shows that the mean field solution captures the instanton
transition between spin up/down states, but fails to reproduce the
damping of coherent oscillations.

To understand the instanton transition in simple terms, we perform
a dRG analysis of the equations. For the present case, where the bath
spins are much slower than the central spin, the natural small parameters
are $\Delta_{\perp}$ and $A_{i}$. Hence we attach a dimensionless
parameter $\epsilon$ to all the terms in Eq.\ref{eq:BathEOM}, in
order to organize the perturbative series around the static bath solution.
This implies that, to order $\epsilon^{0}$, the equations of motion
for the central spin reduce to:
\begin{equation}
\partial_{t}\langle S^{\alpha}\rangle_{0}=\sum_{\mu,\theta=x,y,z}\epsilon_{\mu\alpha\theta}\left(B_{\mu}-\sum_{i=1}^{N}A_{i}^{\mu}m_{i}^{\mu}\right)\langle S_{0}^{\theta}\rangle_{0}\label{eq:Uncorrelated-S}
\end{equation}
where the bath $\partial_{t}\langle I_{i}^{\alpha}\rangle_{0}=0\rightarrow\langle I_{i}^{\alpha}\left(t_{0}\right)\rangle_{0}=m_{i}^{\alpha}$
is static at this order, and $\langle\ldots\rangle_{0}$ indicates
the average value of the unperturbed solution. Notice that due to
the sum over all bath spins, the term $\sum_{i=1}^{N}A_{i}m_{i}^{z}$
is not assumed to be small, and it is present to lowest order in $\epsilon$
(this means that the Overhauser field can be large and contribute
to the fast dynamics of the central spin). The solution to these equations
corresponds to the one found for the static bath case (Eq.\ref{eq:Exact-Dynamics1}),
which will be the starting point of our analysis. The important difference
is that now the equations of motion are non-linear, which allows to
take full advantage of the power of dRG.

To first order in $\epsilon$ the bath becomes dynamical:
\begin{equation}
\partial_{t}\langle I_{i}^{\alpha}\rangle_{1}=\epsilon\sum_{\mu,\theta=x,y,z}\epsilon_{\mu\alpha\theta}\left(\Delta_{\mu}-\langle S^{\mu}\rangle_{0}A_{i}^{\mu}\right)m_{i}^{\theta}
\end{equation}
and the solution displays secular terms, which need to be renormalized.
For example, the longitudinal bath magnetization is given by:
\begin{eqnarray}
\langle I_{i}^{z}\left(t\right)\rangle & = & m_{i}^{z}-\epsilon\left(t-t_{0}\right)\Delta_{\perp}m_{i}^{y}+\mathcal{O}\left(\epsilon^{2}\right)\label{eq:RG0}
\end{eqnarray}
As previously mentioned, the appearance of secular terms is identified
with a breakdown of the perturbative solution for times $\sim1/\epsilon$,
or in this case $\sim1/\epsilon\Delta_{\perp}$. This means that one
can interpret the difference $t-t_{0}$ in Eq.\ref{eq:RG0} as the
distance to a physical cut-off $t_{0}$. However, one can extend this
solution to larger times by making the cut-off dynamical $t_{0}\rightarrow\tau$,
in such a way that the difference $t-\tau\ll1$. Finally, in order
to ensure that the solution does not depend on the arbitrary cut-off,
one must impose:
\begin{eqnarray}
\partial_{\tau}\langle I_{i}^{z}\left(t\right)\rangle & = & 0
\end{eqnarray}
which leads to the next flow equation, to first order in $\epsilon$,
for the boundary condition $m_{i}^{z}\left(\tau\right)$:
\begin{equation}
0=\partial_{\tau}m_{i}^{z}\left(\tau\right)+\epsilon\Delta_{\perp}m_{i}^{y}\left(\tau\right)
\end{equation}
One can derive the flow equations for the other boundary conditions
in a similar way, and this yields:
\begin{eqnarray}
\partial_{\tau}m_{i}^{x}\left(\tau\right) & = & -\epsilon R_{i}\left(\tau\right)m_{i}^{y}\left(\tau\right)\\
\partial_{\tau}m_{i}^{y}\left(\tau\right) & = & \epsilon R_{i}\left(\tau\right)m_{i}^{x}\left(\tau\right)+\epsilon\Delta_{\perp}m_{i}^{z}\left(\tau\right)\\
\partial_{\tau}m_{i}^{z}\left(\tau\right) & = & -\epsilon\Delta_{\perp}m_{i}^{y}\left(\tau\right)
\end{eqnarray}
where
\begin{equation}
R_{i}\left(\tau\right)=A_{i}\frac{B_{\perp}M_{x}\left(\tau\right)+\omega_{z}\left(\tau\right)M_{z}\left(\tau\right)}{\Omega_{\vec{m}}^{2}}\omega_{z}\left(\tau\right)-\Delta_{z}
\end{equation}
and $\omega_{z}=B_{z}-\sum_{i}A_{i}m_{i}^{z}$. It is important to
notice that if $m_{i}^{z}\left(\tau\right)$ is dynamical, the central
spin frequency $\Omega_{\vec{m}}=\sqrt{B_{\perp}^{2}+\left(B_{z}-\sum_{i}A_{i}m_{i}^{z}\left(\tau\right)\right)^{2}}$
will change over time, which is what makes the central spin and the
bath to become resonant at long times, and produces the instanton-like
transition.

\begin{figure}
\includegraphics[scale=0.48]{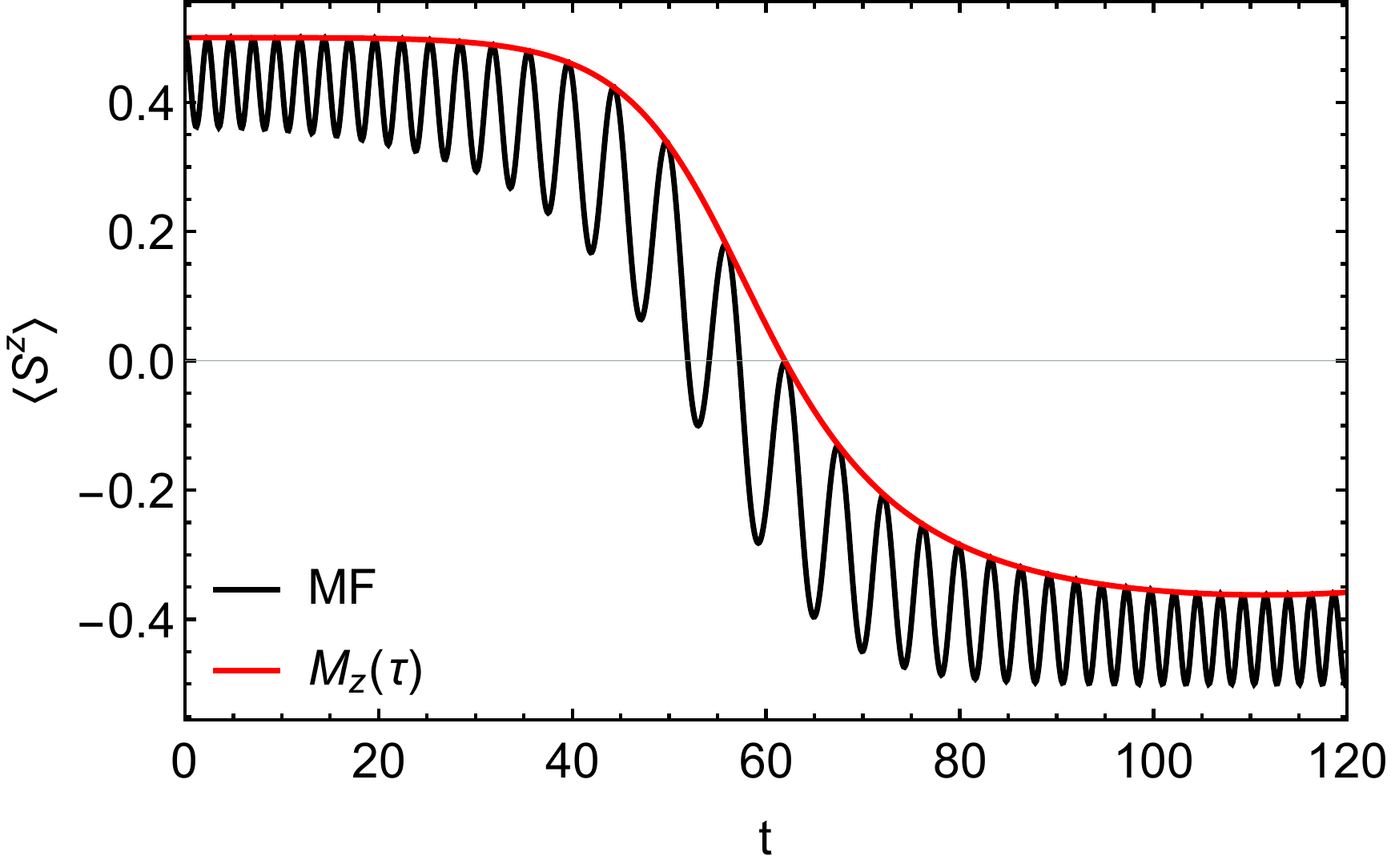}

\caption{\label{fig:FlowEq}Numerical solution of the mean field equations
(black) and the solution for the slow component $M_{z}\left(\tau\right)$
(red) using the flow equation. Parameters:$B_{z}=\Delta_{z}=0$, $\Delta_{\perp}/B_{\perp}=0.03$,
$A/B_{\perp}=0.05$ and $N=100$, with central spin initially up and
a fully polarized bath. The slow component perfectly describes the
instanton transition.}
\end{figure}

As the previous flow equations couple to the boundary conditions for
the central spin $M_{\alpha}\left(\tau\right)$, one must obtain their
flow equation as well. To first order in $\epsilon$, the equation
of motion for the central spin is:
\begin{eqnarray}
\partial_{t}\langle S^{\alpha}\rangle_{1} & = & \sum_{\mu,\theta}\epsilon_{\mu\alpha\theta}\left(B_{\mu}-\sum_{i=1}^{N}A_{i}^{\mu}m_{i}^{\mu}\right)\langle S^{\theta}\rangle_{1}\nonumber \\
 &  & -\sum_{\mu,\theta}\epsilon_{\mu\alpha\theta}\langle S^{\theta}\rangle_{0}\sum_{i=1}^{N}A_{i}^{\mu}\langle I_{i}^{\mu}\rangle_{1}
\end{eqnarray}
and its analytical solution also displays secular terms. These are
due to a resonance with the bath spins, whose longitudinal magnetization
has a fast oscillating component due to the coupling with the central
spin. Once again, assuming that the boundary conditions are dynamical
and the solution is independent of the arbitrary cut-off $\tau$,
the flow equations result in:
\begin{eqnarray}
\partial_{\tau}\log\left(B_{\perp}M_{x}+\omega_{z}M_{z}\right) & = & \epsilon\Delta_{\perp}\frac{\alpha_{y}\omega_{z}}{\Omega_{\vec{m}}^{2}}\label{eq:FlowS1}\\
\partial_{\tau}M_{y} & = & 0\\
\partial_{\tau}\log\left(B_{\perp}M_{z}-\omega_{z}M_{x}\right) & = & \epsilon\Delta_{\perp}\frac{\alpha_{y}\omega_{z}}{\Omega_{\vec{m}}^{2}}
\end{eqnarray}
where we have defined $\alpha_{y}=\sum_{i}A_{i}m_{i}^{y}$. Notice
that the flow of the boundary conditions implies that, even if they
initially vanish, they might become finite over time. The solution
from the flow equations perfectly captures the slow time evolution
that describes the instanton-like transition, as it is shown in Fig.\ref{fig:FlowEq}.
This shows that dRG can be used to separate the dynamics according
to their time scales, and study each of them independently. Furthermore,
the flow equations for the central spin (Eq.\ref{eq:FlowS1}), demonstrate
that the instanton transition is exponentially fast, with exponent
proportional to $\Delta_{\perp}$.

\subsection{Correlation effects}

The mean field equations neglect correlations between the central
spin and the bath spins. That is why they fail to capture the suppression
of coherent oscillations in Fig.\ref{fig:Numerical1}. To show this,
we go one step further and calculate the equation of motion for the
bath-system correlators:
\begin{eqnarray}
\partial_{t}I_{i}^{\beta}S^{\alpha} & = & \left[\epsilon_{x\alpha\theta}B_{\perp}+\epsilon_{z\alpha\theta}\left(B_{z}-\sum_{j\neq i}A_{j}I_{j}^{z}\right)\right]I_{i}^{\beta}S^{\theta}\nonumber \\
 &  & +\left(\epsilon_{x\beta\theta}\Delta_{\perp}+\epsilon_{z\beta\theta}\Delta_{z}\right)I_{i}^{\theta}S^{\alpha}\nonumber \\
 &  & -\frac{A_{i}}{4}\left(\delta_{z,\beta}\epsilon_{z\alpha\theta}S^{\theta}+\delta_{\alpha,z}\epsilon_{z\beta\theta}I_{i}^{\theta}\right)\label{eq:Correlator-1}
\end{eqnarray}
As expected, this equation couples to three-point correlators and
requires a decoupling scheme to find a solution. We have considered
three different decoupling schemes, which are discussed in the Appendix.
The one based on a \emph{Hierarchy of Correlations} is the one that
gives the best results, at least for this model. This decoupling scheme
has been previously discussed\citep{Hierarchy}, and it is based on
the decomposition $\langle I_{i}^{\beta}S^{\alpha}\rangle=\langle I_{i}^{\beta}\rangle\langle S^{\alpha}\rangle+\langle I_{i}^{\beta}S^{\alpha}\rangle^{c}$,
where $\langle\ldots\rangle^{c}$ indicates the correlated part, which
is defined as the difference between the mean field and the exact
value. The reason why this decomposition works better than the other
ones considered is because it organizes the non-linear corrections
in a way that they tend to be always small, compared with the mean
field value.

\begin{figure}
\includegraphics[scale=0.48]{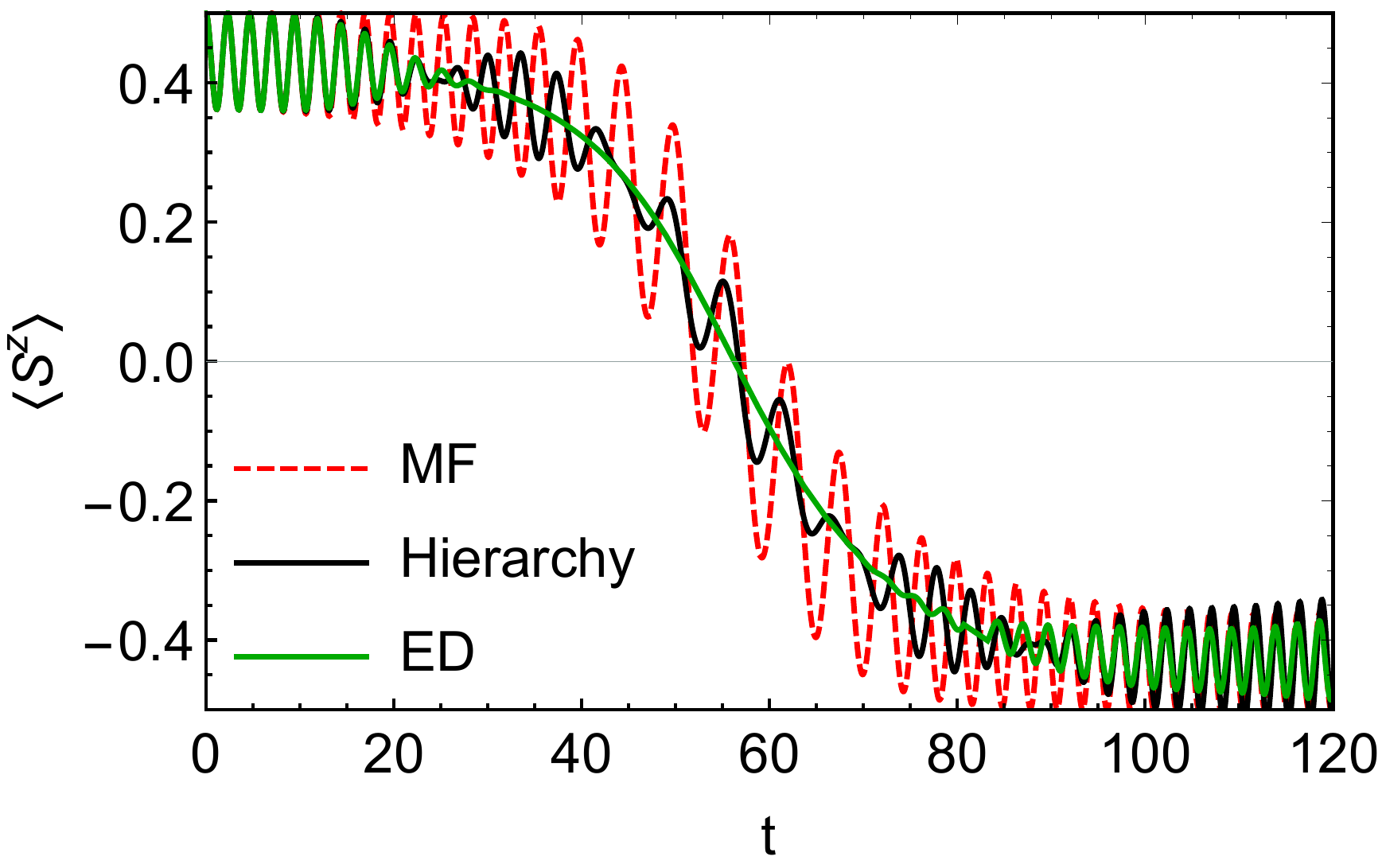}

\caption{\label{fig:Comparison-corr1}Comparison between the mean field solution
(red), the solution including spin-bath correlations (black) and the
exact simulation (green) for the same parameters as Fig.\ref{fig:Numerical1}.}
\end{figure}

The separation into correlated and uncorrelated parts leads to the
next final equation of motion for the correlated parts:
\begin{eqnarray}
\partial_{t}\langle I_{i}^{\beta}S^{\alpha}\rangle^{c} & \simeq & \epsilon_{z\alpha\mu}\left(B_{z}-\sum_{j\neq i}^{N}A_{j}\langle I_{j}^{z}\rangle\right)\langle I_{i}^{\beta}S^{\mu}\rangle^{c}\nonumber \\
 &  & +\epsilon_{x\alpha\mu}B_{\perp}\langle I_{i}^{\beta}S^{\mu}\rangle^{c}\nonumber \\
 &  & -\epsilon_{z\alpha\mu}A_{i}\langle S^{\mu}\rangle\left(\frac{\delta_{z,\beta}}{4}-\langle I_{i}^{\beta}\rangle\langle I_{i}^{z}\rangle\right)\nonumber \\
 &  & -\epsilon_{z\beta\nu}A_{i}\langle I_{i}^{\nu}\rangle\left(\frac{\delta_{z,\alpha}}{4}-\langle S^{\alpha}\rangle\langle S^{z}\rangle\right)\label{eq:Correlator-2}
\end{eqnarray}
In order to obtain Eq.\ref{eq:Correlator-2} we have assumed that
correlated parts are small (at least for short time), neglected bath-bath
correlators and terms proportional to $\vec{\Delta}$. These terms
can be neglected because they produce slower dynamics, however, finding
the solution in their presence is not difficult (detailed derivation
in the Appendix). This equation must be numerically solved simultaneously
with the equations for the bath and central spin:
\begin{eqnarray}
\partial_{t}\langle S^{\alpha}\rangle & \simeq & \sum_{\mu,\theta}\epsilon_{\mu\alpha\theta}\left(B_{\mu}-\sum_{i=1}^{N}A_{i}^{\mu}\langle I_{i}^{\mu}\rangle\right)\langle S^{\theta}\rangle\label{eq:Correlator-S}\\
 &  & -\sum_{\mu,\theta}\epsilon_{\mu\alpha\theta}\sum_{i=1}^{N}A_{i}^{\mu}\langle I_{i}^{\mu}S^{\theta}\rangle^{c}\nonumber \\
\partial_{t}\langle I_{i}^{\alpha}\rangle & \simeq & \sum_{\mu,\theta}\epsilon_{\mu\alpha\theta}\left(\Delta_{\mu}-A_{i}^{\mu}\langle S^{\mu}\rangle\right)\langle I_{i}^{\theta}\rangle\label{eq:Correlator-I}
\end{eqnarray}
where we have also neglected the slow terms in the equation of motion
for the bath spin, which are proportional to the correlated part.
The numerical solution of the equations of motion, including spin-bath
correlators is shown in Fig.\ref{fig:Comparison-corr1}. It shows
that the addition of lowest order corrections, due to system-bath
correlations, allows to capture the suppression of coherent oscillations.
At longer time-scales other processes involving many-body correlations
take over, but the qualitative behavior of the magnetization is correctly
captured. This can be fixed by including extra terms in the equation
of motion for the system-bath correlations.
\\

As we know that Eq.\ref{eq:Correlator-2} correctly captures the main
features of the dynamics, we now analyze the equations using dRG,
to unravel the role of correlations between spins. Importantly, this
time it will lead to flow equations for the quantum correlations between
the spins, and demonstrate which contributions are crucial as time
evolves, even for initial product states where correlations vanish.

For the perturbative solution we expand again, in powers of $\epsilon$,
the equations of motion (Eqs.\ref{eq:Correlator-2}, \ref{eq:Correlator-S}
and \ref{eq:Correlator-I}). For simplicity, we also assume that correlated
parts are small and attach a factor $\epsilon$ to them in Eqs.\ref{eq:Correlator-S}
and \ref{eq:Correlator-I}. This will indeed be the case at short
time, if the system initially is uncorrelated with the bath. However,
it neglects an important backreaction between fluctuations and mean
field values that will affect the frequency of the oscillations. Because
correlated parts are proportional to $\epsilon$, to lowest order
the equations of motion still are the mean field equations previously
solved, with the addition of the lowest order equation of motion for
the correlated part:
\begin{eqnarray}
\partial_{t}\langle I_{i}^{\beta}S^{\alpha}\rangle_{0}^{c} & \simeq & \left(\epsilon_{z\alpha\mu}\omega_{z}+\epsilon_{x\alpha\mu}B_{\perp}\right)\langle I_{i}^{\beta}S^{\mu}\rangle_{0}^{c}
\end{eqnarray}
This equation is analogous to the one for the central spin (Eq.\ref{eq:Uncorrelated-S}),
with just different boundary condition. To first order in $\epsilon$
the bath equation of motion is still unchanged with respect to the
mean field case, as correlation terms are of order $\epsilon^{2}$.
This is not the case for the central spin, where correlated and uncorrelated
parts couple in the equation of motion:
\begin{eqnarray}
\partial_{t}\langle S^{\alpha}\rangle_{1} & = & \sum_{\mu,\theta}\epsilon_{\mu\alpha\theta}\left(B_{\mu}-\sum_{i=1}^{N}A_{i}^{\mu}m_{i}^{\mu}\right)\langle S^{\theta}\rangle_{1}\\
 &  & -\sum_{\mu,\theta}\epsilon_{\mu\alpha\theta}\langle S^{\theta}\rangle_{0}\sum_{i=1}^{N}A_{i}^{\mu}\langle I_{i}^{\mu}\rangle_{1}\nonumber \\
 &  & -\epsilon\sum_{\mu,\theta}\epsilon_{\mu\alpha\theta}\sum_{i=1}^{N}A_{i}^{\mu}\langle I_{i}^{\mu}S^{\theta}\rangle_{0}^{c}\nonumber 
\end{eqnarray}
where the last line corresponds to the lowest order solution for the
correlated part. The solution displays once again secular terms, however,
the addition of correlations produces corrections to the flow equations
obtained in the mean field case (Eq.\ref{eq:FlowS1}). They are now
given by:
\begin{eqnarray}
\partial_{\tau}\log\left(B_{\perp}M_{x}+M_{z}\omega_{z}\right) & = & \epsilon\Delta_{\perp}\omega_{z}\frac{\eta_{y}}{\Omega_{\vec{m}}^{2}}\label{eq:FlowCorr-S-1}\\
\partial_{\tau}\left(B_{\perp}M_{z}-\omega_{z}M_{x}\right) & = & \epsilon\omega_{z}\Delta_{\perp}\eta_{y}\frac{B_{\perp}M_{z}-\omega_{z}M_{x}}{\Omega_{\vec{m}}^{2}}\nonumber \\
 &  & +\epsilon\omega_{z}a_{zy}\nonumber \\
\partial_{\tau}M_{y} & = & \epsilon\omega_{z}\frac{a_{zx}\omega_{z}-B_{\perp}a_{zz}}{\Omega_{\vec{m}}^{2}}\nonumber 
\end{eqnarray}
where $a_{\alpha\beta}=\sum_{i}A_{i}c_{i}^{\alpha\beta}$, and $c_{i}^{\alpha\beta}\left(t_{0}\right)$
is the initial condition for the correlated part $\langle I_{i}^{\alpha}S^{\beta}\rangle^{c}$.
The most important change with respect to the mean field case (Eq.\ref{eq:FlowS1})
is that now $M_{y}$ can flow, and that the boundary condition for
the correlated part also affects the longitudinal and transverse magnetization.
Furthermore, assuming that correlations do not develop over time ($a_{\alpha\beta}\left(\tau\right)=0\ \forall\ \tau$),
one recovers the mean field flow equations.

\begin{figure}
\includegraphics[scale=0.48]{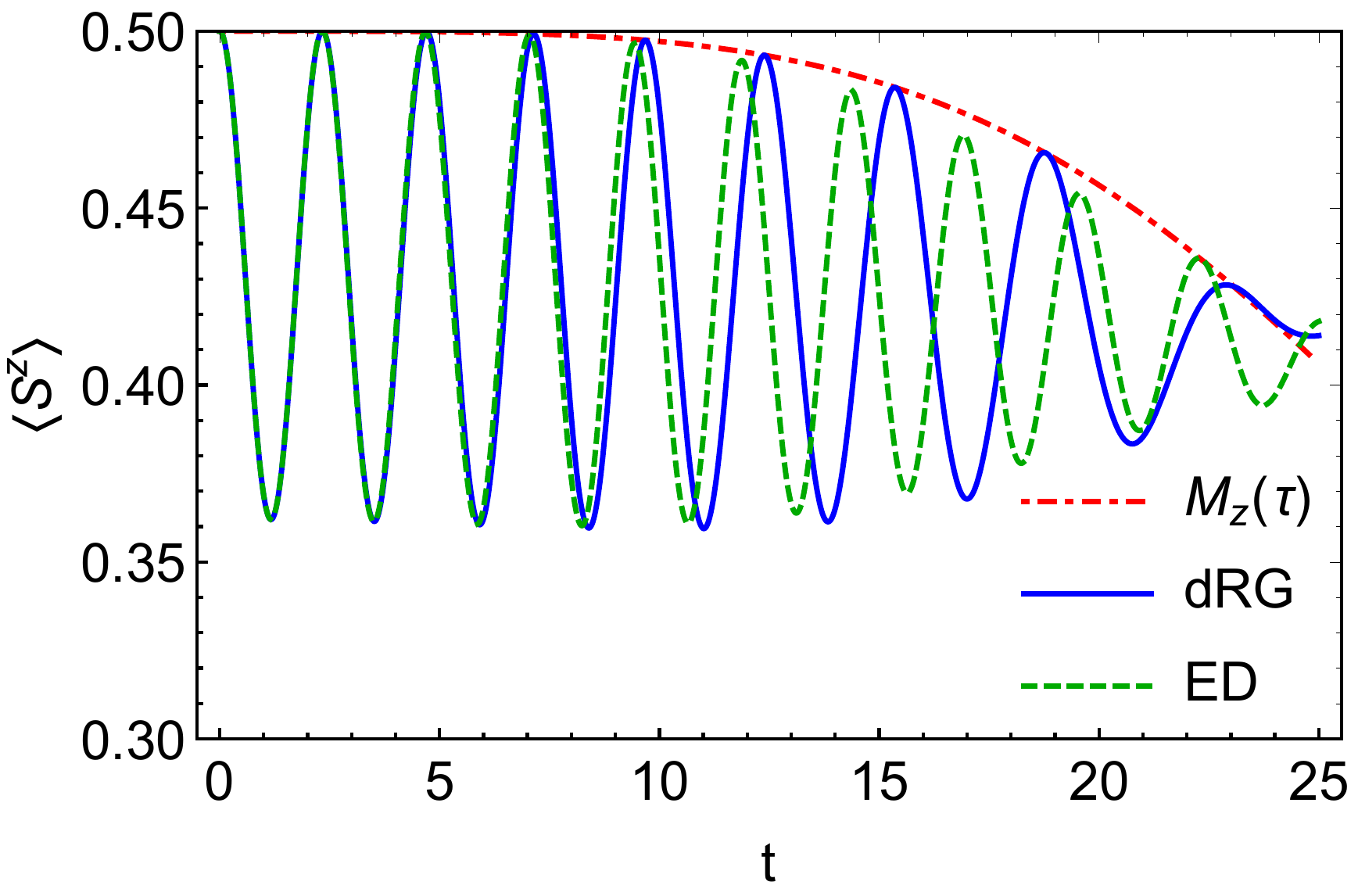}

\caption{\label{fig:Comparison2-1}Comparison between the exact dynamics (dashed,
green) and the lowest order solution using dRG (blue) for the same
parameters as Fig.\ref{fig:Numerical1}. The red dot-dashed line shows
$M_{z}\left(\tau\right)$, which plays the role of the envelope function
for the faster oscillations. The lowest order solution correctly captures
the suppression of oscillations, but the frequency is shifted because
backreaction between the mean field solution and the correlations
has been neglected at this order of dRG.}
\end{figure}

As previously mentioned, the addition of correlated parts implies
that now their boundary conditions $c_{i}^{\alpha\beta}$ will be
renormalized over time, if the solution to the equation of motion
to first order in $\epsilon$ has secular terms:
\begin{eqnarray}
\partial_{t}\langle I_{i}^{\beta}S^{\alpha}\rangle_{1}^{c} & \simeq & \left(\epsilon_{z\alpha\mu}\omega_{z}+\epsilon_{x\alpha\mu}B_{\perp}\right)\langle I_{i}^{\beta}S^{\mu}\rangle_{1}^{c}\\
 &  & -\epsilon_{z\alpha\mu}\langle I_{i}^{\beta}S^{\mu}\rangle_{0}^{c}\sum_{j\neq i}^{N}A_{j}\langle I_{j}^{z}\rangle_{1}\nonumber \\
 &  & -\epsilon_{z\alpha\mu}\epsilon A_{i}\langle S^{\mu}\rangle_{0}\left(\frac{\delta_{z,\beta}}{4}-m_{i}^{\beta}m_{i}^{z}\right)\nonumber \\
 &  & -\epsilon_{z\beta\nu}\epsilon A_{i}m_{i}^{\nu}\left(\frac{\delta_{z,\alpha}}{4}-\langle S^{\alpha}\rangle_{0}\langle S^{z}\rangle_{0}\right)\nonumber 
\end{eqnarray}
This is the case, and the flow equations for the initial correlations
are given by:
\begin{eqnarray}
\partial_{\tau}\log\left(B_{\perp}a_{zx}+a_{zz}\omega_{z}\right) & = & \epsilon\Delta_{\perp}\omega_{z}\frac{\eta_{y}}{\Omega_{\vec{m}}^{2}}\label{eq:FlowCorr-C-1}\\
\partial_{\tau}\left(B_{\perp}a_{zz}-a_{zx}\omega_{z}\right) & = & \epsilon\Delta_{\perp}\omega_{z}\frac{\eta_{y}}{\Omega_{\vec{m}}^{2}}\left(B_{\perp}a_{zz}-a_{zx}\omega_{z}\right)\nonumber \\
 &  & +\epsilon M_{y}\omega_{z}\xi_{z}\\
\partial_{\tau}a_{zy} & = & -\epsilon\frac{\xi_{z}\omega_{z}}{\Omega^{2}}\left(B_{\perp}M_{z}-\omega_{z}M_{x}\right)
\end{eqnarray}
where we have defined $\xi_{z}=\sum_{i=1}^{N}A_{i}^{2}\left(\frac{1}{4}-\left(m_{i}^{z}\right)^{2}\right)$.
The suppression of oscillations is obtained due to the $\xi_{z}$
function, which captures the precession of the bath magnetization
away from the longitudinal axis. Also, even for vanishing initial
correlations, $a_{zy}\left(\tau\right)$ becomes non-negligible over
time, as it is proportional to $\xi_{z}$ and to the central spin
magnetization. Fig.\ref{fig:Comparison2-1} shows a comparison between
the exact dynamics, the one numerically obtained by the hierarchy
of correlations decoupling and its lowest order approximation obtained
from dRG.
It can be seen that the first order approximation provides good agreement
for the amplitude renormalization, however there is a frequency shift
with respect to the exact solution. The reason for this discrepancy
is that the backreaction between uncorrelated and correlated parts
was neglected to lowest order, which only holds for short times (In
Fig.\ref{fig:Comparison-corr1} this is included and one can see that
the frequency of the exact solution and that of the numerical solution
of Eq.\ref{eq:Correlator-2} coincide).

\section{Conclusions}

We have shown that it is possible to obtain good approximations for
the dynamics of strongly correlated systems, such as the central spin
model, by numerical and analytical methods.

In the first part we have discussed the differences between an external
magnetic field and a static spin bath by calculating the exact solution
of the model. Then, we have demonstrated that when the environment
is in an excited state, destructive interference between different
quantum states results in a suppression of the coherent oscillations,
which can be characterized by a Gaussian spectral function. This is
not a decoherence process however, as entanglement between the two
systems is not created, and it can be reversed with spin echo. Importantly,
it is interesting that the suppression is highly dependent on the
Zeeman splitting of the central spin. This property can be used to
characterize some of the properties of the environment.

In the second part we have included a transverse field acting on the
bath spins, to switch-on their dynamics. It is shown that non-perturbative
effects can be important after a short time. We have numerically solved
the model finding that, a separation of the equations of motion into
mean field and quantum fluctuations, provides good agreement with
the exact dynamics, once the lowest order fluctuations are added.
The main features of this model are: Amplitude modulation, and the
suppression of coherent oscillations due to entanglement with the
environment. On the one hand, the amplitude modulation, which produces
an instanton-like transition, is well captured at the mean field level.
On the other hand, the suppression of coherent oscillations requires
the quantum fluctuations to be included. Importantly, the suppression
of oscillations in the case of a dynamical bath is linked with the
formation of entanglement with the spin bath, unlike in the case of
a static bath, and cannot generally be removed by spin echo techniques
(the spin bath is precessing under a different magnetic field $\vec{\Delta}$).

Finally, we have shown that the equations of motion for the model
can be analyzed using dynamical Renormalization Group techniques.
The advantages of this technique are several: i) It provides an analytical
approach to the highly complex numerical solutions, ii) Provides non-perturbative
results, and iii) it can eliminate secular terms, even when they are
present in the full numerical solution. It is also interesting that
when quantum fluctuations are included, one finds non-perturbative
expressions for the entanglement between system and environment, which
can be useful for state preparation in experiments with many particles.

Our results can be easily applied to study the dynamics of other models
of qubits interacting with surrounding localized modes, which is important
for the design of quantum computers, as these are expected to dominate
$T_{2}$ at low temperatures\citep{KaneQC,StampNature,TheoryBathSpin2000,AdjustableSpinBath}.
Furthermore, we expect this approach to be able to characterize the
decoherence rates in cases where simple Markovian solutions can fail.
\begin{acknowledgments}
We thank T. Cox, P.C.E. Stamp, G. Platero, T. Staubert and S. Kehrein
for insightful discussions. This work was supported by the Spanish
Ministry of Economy and Competitiveness through Grant MAT2014-58241-P,
Grant MAT2017-86717-P and the Juan de la Cierva program. We also acknowledge
support from the CSIC Research Platform on Quantum Technologies PTI-001.
\end{acknowledgments}

\bibliographystyle{phaip}
\bibliography{Bibliography}

\begin{widetext}

\appendix

\section{Exact solution for a static bath}

The dynamics for a central spin, longitudinally coupled to a static
bath of spins can be exactly solved. Starting from the Hamiltonian
in Eq.\ref{eq:H}, one can make use of the basis of eigenstates for
the case with $B_{\perp}=\Delta_{\perp}=0$, given by $|S,M;\vec{P},\vec{m}\rangle$,
where $S$ indicates the spin value of the central spin, $M\in\left[-S,S\right]$
its projection onto the z-axis, $\vec{P}=\left(P_{1},P_{2},\ldots,P_{N}\right)$
is the spin value of the different bath spins, and $\vec{m}=\left(m_{1},m_{2},\ldots,m_{N}\right)$
their projection $m_{i}\in\left[-P_{i},P_{i}\right]$. The Hamiltonian
in this basis is given by:
\begin{eqnarray}
H_{0} & = & \sum_{S,\vec{P}}\sum_{M,\vec{m}}\left(-B_{z}M-\Delta_{z}\sum_{i}m_{i}+M\sum_{i}A_{i}m_{i}\right)X_{\vec{m},\vec{m}}^{M,M}\\
V_{S} & = & -\frac{B_{\perp}}{2}\left(S^{+}+S^{-}\right)=-\frac{B_{\perp}}{2}\sum_{S,\vec{P}}\sum_{M,\vec{m}}\gamma_{S,M}\left(X_{\vec{m},\vec{m}}^{M+1,M}+X_{\vec{m},\vec{m}}^{M,M+1}\right)\\
V_{B} & = & -\frac{\Delta_{\perp}}{2}\sum_{i}\left(I_{i}^{+}+I_{i}^{-}\right)=-\frac{\Delta_{\perp}}{2}\sum_{i}\sum_{S,\vec{P}}\sum_{M,\vec{m}}\gamma_{P_{i},m_{i}}\left(X_{\vec{m}+1_{i},\vec{m}}^{M,M}+X_{\vec{m},\vec{m}+1_{i}}^{M,M}\right)
\end{eqnarray}
where $X_{\vec{m},\vec{m}^{\prime}}^{M,M^{\prime}}=|S,M;\vec{P},\vec{m}\rangle\langle S^{\prime},M^{\prime};\vec{P}^{\prime},\vec{m}^{\prime}|$
are the Hubbard operators, $\gamma_{P_{i},m_{i}}=\sqrt{P_{i}\left(P_{i}+1\right)-m_{i}\left(m_{i}+1\right)}$
and $\vec{m}\pm1_{i}$ indicates that for the spin configuration $\vec{m}$,
the spin projection $m_{i}\rightarrow m_{i}\pm1$, leaving all the
other $m_{j}$, for all $j\neq i$, unchanged.

For the present case with $S=1/2$, we can easily diagonalize $H_{0}+V_{S}$,
because the Hilbert space factorizes in different bath configurations
$\vec{m}$. The equations of motion for the different projection operators
$X_{\vec{m},\vec{m}}^{\pm,\pm}$ are obtained in the Heisenberg picture
using the Heisenberg equation of motion $\partial_{t}\hat{O}=i\left[H,\hat{O}\right]$:
\begin{eqnarray}
\partial_{t}X_{\vec{m},\vec{m}^{\prime}}^{+,+} & = & i\left(\omega_{\vec{m}}^{+}-\omega_{\vec{m}^{\prime}}^{+}\right)X_{\vec{m},\vec{m}^{\prime}}^{+,+}-i\frac{B_{\perp}}{2}\left(X_{\vec{m},\vec{m}^{\prime}}^{-,+}-X_{\vec{m},\vec{m}^{\prime}}^{+,-}\right)\\
\partial_{t}X_{\vec{m},\vec{m}^{\prime}}^{-,-} & = & i\left(\omega_{\vec{m}}^{-}-\omega_{\vec{m}^{\prime}}^{-}\right)X_{\vec{m},\vec{m}^{\prime}}^{-,-}-i\frac{B_{\perp}}{2}\left(X_{\vec{m},\vec{m}^{\prime}}^{+,-}-X_{\vec{m},\vec{m}^{\prime}}^{-,+}\right)\\
\partial_{t}X_{\vec{m},\vec{m}^{\prime}}^{+,-} & = & i\left(\omega_{\vec{m}}^{+}-\omega_{\vec{m}^{\prime}}^{-}\right)X_{\vec{m},\vec{m}^{\prime}}^{+,-}-i\frac{B_{\perp}}{2}\left(X_{\vec{m},\vec{m}^{\prime}}^{-,-}-X_{\vec{m},\vec{m}^{\prime}}^{+,+}\right)\\
\partial_{t}X_{\vec{m},\vec{m}^{\prime}}^{-,+} & = & i\left(\omega_{\vec{m}}^{-}-\omega_{\vec{m}^{\prime}}^{+}\right)X_{\vec{m},\vec{m}^{\prime}}^{-,+}-i\frac{B_{\perp}}{2}\left(X_{\vec{m},\vec{m}^{\prime}}^{+,+}-X_{\vec{m},\vec{m}^{\prime}}^{-,-}\right)
\end{eqnarray}
with $\omega_{\vec{m}}^{M}=-M\left(B_{z}-\vec{A}\cdot\vec{m}\right)-\Delta_{z}\sum_{i}m_{i}$.
The solutions can be directly obtained; however as we are interested
in the central spin dynamics, the solution for the time evolution
of the different central spin operators is even simpler (because they
are diagonal in the bath indices):
\begin{eqnarray}
S_{\vec{m}}^{x}\left(t\right) & = & S_{\vec{m}}^{x}\frac{B_{\perp}^{2}+\left(B_{z}-\vec{A}\cdot\vec{m}\right)^{2}\cos\left(\Omega_{\vec{m}}t\right)}{\Omega_{\vec{m}}^{2}}+S_{\vec{m}}^{y}\frac{B_{z}-\vec{A}\cdot\vec{m}}{\Omega_{\vec{m}}}\sin\left(\Omega_{\vec{m}}t\right)+S_{\vec{m}}^{z}\frac{B_{\perp}\left(B_{z}-\vec{A}\cdot\vec{m}\right)}{\Omega_{\vec{m}}^{2}}\left[1-\cos\left(\Omega_{\vec{m}}t\right)\right]\\
S_{\vec{m}}^{y}\left(t\right) & = & S_{\vec{m}}^{y}\cos\left(\Omega_{\vec{m}}t\right)+\frac{S_{\vec{m}}^{z}B_{\perp}-S_{\vec{m}}^{x}\left(B_{z}-\vec{A}\cdot\vec{m}\right)}{\Omega_{\vec{m}}}\sin\left(\Omega_{\vec{m}}t\right)\\
S_{\vec{m}}^{z}\left(t\right) & = & S_{\vec{m}}^{z}\frac{\left(B_{z}-\vec{A}\cdot\vec{m}\right)^{2}+B_{\perp}^{2}\cos\left(\Omega_{\vec{m}}t\right)}{\Omega_{\vec{m}}^{2}}-S_{\vec{m}}^{y}B_{\perp}\frac{\sin\left(\Omega_{\vec{m}}t\right)}{\Omega_{\vec{m}}}+S_{\vec{m}}^{x}B_{\perp}\left(B_{z}-\vec{A}\cdot\vec{m}\right)\frac{1-\cos\left(\Omega_{\vec{m}}t\right)}{\Omega_{\vec{m}}^{2}}
\end{eqnarray}
where we have just rewritten the equations of motion in the basis
of spin operators for a given bath configuration: 
\begin{eqnarray*}
S_{\vec{m}}^{x}\left(t\right) & = & \frac{1}{2}\left(X_{\vec{m},\vec{m}}^{+,-}\left(t\right)+X_{\vec{m},\vec{m}}^{-,+}\left(t\right)\right)\\
S_{\vec{m}}^{y}\left(t\right) & = & -\frac{i}{2}\left(X_{\vec{m},\vec{m}}^{+,-}\left(t\right)-X_{\vec{m},\vec{m}}^{-,+}\left(t\right)\right)\\
S_{\vec{m}}^{z}\left(t\right) & = & \frac{1}{2}\left(X_{\vec{m},\vec{m}}^{+,+}\left(t\right)-X_{\vec{m},\vec{m}}^{-,-}\left(t\right)\right)
\end{eqnarray*}
In the previous expressions we have defined the frequency for a given
bath configuration $\Omega_{\vec{m}}=\sqrt{B_{\perp}^{2}+\left(B_{z}-\vec{A}\cdot\vec{m}\right)^{2}}$.
The total time evolution for the magnetization is given by $\vec{S}\left(t\right)=\sum_{\vec{P},\vec{m}}\vec{S}_{\vec{m}}\left(t\right)$,
which requires to sum over all spin bath configurations. In the next
Appendix it is shown how one can easily do this.

\section{Sum over polarization groups}

The exact expression for the magnetization of the central spin requires
to sum over all spin bath configurations, and depending on the initial
state, the results can be quite different. First of all, it is useful
to consider a situation where the initial state for the total system
is a product state, as typically an experiment can control the central
spin/qubit, but not the environmental degrees of freedom:
\begin{eqnarray}
\rho\left(t_{0}\right) & = & \rho_{S}\otimes\rho_{B}
\end{eqnarray}
This is only useful for our discussion, but not required for the derivation.
Now it is important to realize that two opposite situations can happen:
The environment it is either in its ground state (i.e., the temperature
is low enough that just the lowest energy states are occupied), or
its in a high temperature state (i.e., thermal activation equally
occupies all the bath modes). In the first case, the sum over bath
configurations is dominated by a single term and the sum does not
need to be calculated. This implies that the central spin dynamics
will only contain a single frequency and the dynamics can be easily
understood. In the second case the sum has a huge number of frequencies,
and the total summation can be cumbersome. In this case, some approximate
method to simplify the sum would be desirable. In this case it is
useful to consider a density of states (DOS) $J\left(\alpha\right)$
such that:
\begin{eqnarray}
\vec{S}\left(t\right) & = & \sum_{\vec{P},\vec{m}}\vec{S}_{\vec{m}}\left(t\right)=\int_{-\infty}^{\infty}d\alpha\vec{S}\left(t,\alpha\right)J\left(\alpha\right)\\
J\left(\alpha\right) & \equiv & \sum_{\vec{m}}g_{\vec{m}}\delta\left(\alpha-B_{z}+\vec{A}\cdot\vec{m}\right)
\end{eqnarray}
where $g_{\vec{m}}$ accounts for the degeneracy of each polarization
group configuration. Using a Fourier transform one can rewrite the
DOS as:
\begin{eqnarray}
J\left(\alpha\right) & = & \int_{-\infty}^{\infty}\frac{d\epsilon}{2\pi}\sum_{\vec{m}}g_{\vec{m}}e^{i\epsilon\left(\alpha-B_{z}+\vec{A}\cdot\vec{m}\right)}\\
 & = & \int_{-\infty}^{\infty}\frac{d\epsilon}{2\pi}e^{i\epsilon\left(\alpha-B_{z}\right)}\prod_{i=1}^{N}\sum_{m_{i}=-P_{i}}^{P_{i}}g_{m_{i}}e^{i\epsilon A_{i}m_{i}}\nonumber 
\end{eqnarray}
where $g_{m_{i}}=P_{i}-\left|m_{i}\right|+1$. At this point, depending
on the spin value $P_{i}$, one can approximate the sum as an integral
if $P_{i}\gg1/2$. If this is not the case one can directly calculate
the sum, but we will consider the case $P_{i}\gg1$, because it leads
to more compact expressions and is valid in many cases. Finally, one
can calculate the large product $\prod_{i=1}^{N}$ using a stationary
phase approximation, which leads to the final expression for the normalized
DOS:
\begin{equation}
J\left(\alpha\right)=\frac{e^{-\frac{\left(\alpha-B_{z}\right)^{2}}{2\sigma^{2}}}}{\sqrt{2\pi\sigma^{2}}},\ \sigma\equiv\sqrt{\frac{1}{6}\sum_{i}A_{i}^{2}P_{i}^{2}\frac{P_{i}+4}{P_{i}+2}}
\end{equation}
This indicates that for the case of broadening larger than the hyperfine
splitting, the effect of the bath is similar to an statistical average
over a bias field with Gaussian fluctuations. Furthermore, the Gaussian
broadens as $N^{1/2}$ with the number of bath spins $N$, and also
linearly with $A_{i}$ and $P_{i}$. This implies that the final expression
for the central spin magnetization, is given by:
\begin{equation}
\vec{S}\left(t\right)=\int_{-\infty}^{+\infty}\vec{S}_{\alpha}\left(t\right)J\left(\alpha\right)d\alpha
\end{equation}
with $\Omega_{\alpha}=\sqrt{B_{\perp}^{2}+\alpha^{2}}$. The expressions
for the time evolution of the different components of the central
spin magnetization become:
\begin{eqnarray}
S_{\alpha}^{x}\left(t\right) & = & S^{x}\frac{B_{x}^{2}+\alpha^{2}\cos\left(\Omega_{\alpha}t\right)}{\Omega_{\alpha}^{2}}+S^{y}\frac{\alpha}{\Omega_{\alpha}}\sin\left(\Omega_{\alpha}t\right)+S^{z}\frac{B_{\perp}\alpha}{\Omega_{\alpha}^{2}}\left[1-\cos\left(\Omega_{\alpha}t\right)\right]\\
S_{\alpha}^{y}\left(t\right) & = & S^{y}\cos\left(\Omega_{\alpha}t\right)+\frac{S^{z}B_{x}-S^{x}\alpha}{\Omega_{\alpha}}\sin\left(\Omega_{\alpha}t\right)\\
S_{\alpha}^{z}\left(t\right) & = & S^{z}\frac{\alpha^{2}+B_{\perp}^{2}\cos\left(\Omega_{\alpha}t\right)}{\Omega_{\alpha}^{2}}-B_{\perp}S^{y}\frac{\sin\left(\Omega_{\alpha}t\right)}{\Omega_{\alpha}}+B_{\perp}\alpha S^{x}\frac{1-\cos\left(\Omega_{\alpha}t\right)}{\Omega_{\alpha}^{2}}
\end{eqnarray}
For intermediate temperature regimes, one can consider a thermal distribution
for the occupation of the different hyperfine levels, but at high
enough $T$, its value is equal for all of them. These expressions
are valid up to some long recurrence time $\tau_{P}\sim A_{i}^{-1}$
for $A_{i}$ small, but the integral with the Gaussian function produces
a behavior that emulates decoherence, as it is shown in Fig.\ref{fig:Comparison0}.
Furthermore, the recurrence time will not be typically captured in
experiments, because at long times, one expects that phonons and other
delocalized modes will take over. Hence this result should be a very
good approximation to describe the short time dynamics under the influence
of an almost static spin bath.

Interestingly, when the standard deviation $\sigma$ is of the order
of the central spin splitting $\sigma/B_{z}\sim1$, coherent oscillations
are only weakly damped, indicating that this could be a good regime
to operate with the qubit. It also would allow to experimentally access
to information about the bath, by sweeping over $B_{z}$ and monitoring
the dynamics (to estimate the number of modes, their spin or the coupling
strength). Finally, the assumption of high temperature in the bath
is not strictly necessary, and this result is valid for any case where
the bath state is a large superposition of different configurations,
even at $T=0$. Notice that the resulting suppression of the coherent
oscillations is not due to a disorder bias average (it also happens
for the case $A_{i}=A$), but it is a consequence of the bath being
a quantum system which can be in a superposition state, and the different
phases of the different configurations interfere destructively. This
indicates that the spin bath cannot be simply thought as a classical
magnetic field.

\section{Perturbative dynamics}

When the transverse field acting on the bath is turned on, different
spin configurations couple, and for large systems, the exact calculation
becomes cumbersome. To estimate the effect of the transverse field
acting on the bath spins, it is useful to transform to the basis of
eigenstates for $H_{S}=H_{0}+V_{S}$. As the Hamiltonian is diagonal
in the bath configurations $\vec{m}$, the diagonalization is simple
and leads to:
\begin{eqnarray}
H_{S} & = & \sum_{\vec{P},\vec{m}}H_{S}\left(\vec{m}\right)\\
H_{S}\left(\vec{m}\right) & = & \left(\begin{array}{cc}
-\Delta_{z}\sum_{i}m_{i}-\frac{B_{z}}{2}+\frac{\vec{A}}{2}\cdot\vec{m} & -\frac{B_{\perp}}{2}\\
-\frac{B_{\perp}}{2} & -\Delta_{z}\sum_{i}m_{i}+\frac{B_{z}}{2}-\frac{\vec{A}}{2}\cdot\vec{m}
\end{array}\right)\label{eq:Ham2x2}
\end{eqnarray}
with eigenvalues
\begin{equation}
E_{\vec{m}}^{M}=-\Delta_{z}\sum_{i}m_{i}+M\sqrt{B_{\perp}^{2}+\left(B_{z}-\vec{A}\cdot\vec{m}\right)^{2}}
\end{equation}
In this basis the full Hamiltonian becomes:
\begin{eqnarray}
H & = & \sum_{\vec{P},\vec{m},M}\left[E_{\vec{m}}^{M}X_{\vec{m},\vec{m}}^{M,M}-\frac{\Delta_{\perp}}{2}\sum_{i}\gamma_{P_{i},m_{i}}\left(X_{\vec{m}+1_{i},\vec{m}}^{M,M}+X_{\vec{m},\vec{m}+1_{i}}^{M,M}\right)\right]
\end{eqnarray}
where now the index $M=\pm1/2$ refers to the eigenstates and eigenvalues
of the matrix in Eq.\ref{eq:Ham2x2}. The dynamics, once the transverse
bath operator is present, is obtained from the time-dependent Schrödinger
equation:
\begin{equation}
i\partial_{t}|\Psi\left(t\right)\rangle=\left(H_{0}+V_{B}\right)|\Psi\left(t\right)\rangle
\end{equation}
Using a decomposition for a general state $|\Psi\left(t\right)\rangle$
in terms of the unperturbed eigenstates $H_{0}|M,\vec{m}\rangle=E_{\vec{m}}^{M}|M,\vec{m}\rangle$,
leads to (we are ignoring the indices $\vec{P}$ and $S$ because
they do not play an important role. However they must be added at
the end of the calculation):
\begin{equation}
|\Psi\left(t\right)\rangle=\sum_{M,\vec{m}}c_{M,\vec{m}}\left(t\right)e^{-itE_{\vec{m}}^{M}}|M,\vec{m}\rangle
\end{equation}
We can now rewrite the time-dependent Schrödinger equation, by multiplying
by $\langle M^{\prime},\vec{m}^{\prime}|$ from the left, as follows:
\begin{equation}
\dot{c}_{M,\vec{m}}\left(t\right)=-i\sum_{\vec{m}^{\prime}}c_{M,\vec{m}^{\prime}}\left(t\right)e^{it\left(E_{\vec{m}}^{M}-E_{\vec{m}^{\prime}}^{M}\right)}\langle\vec{m}|V_{B}|\vec{m}^{\prime}\rangle
\end{equation}
where we have used that the matrix elements of $V_{B}$ only couple
different bath configurations. As in this equation all the different
bath configurations couple, when the bath is large, it must be truncated.
For this, we consider a powers expansion:
\begin{equation}
c_{M,\vec{m}}\left(t\right)=c_{M,\vec{m}}^{\left(0\right)}+\Delta_{\perp}c_{M,\vec{m}}^{\left(1\right)}\left(t\right)+\Delta_{\perp}^{2}c_{M,\vec{m}}^{\left(2\right)}\left(t\right)+\ldots
\end{equation}
Similarly, the calculation of the matrix elements yields:
\begin{eqnarray}
\langle M,\vec{m}|V_{B}|M^{\prime},\vec{m}^{\prime}\rangle & = & -\frac{\Delta_{\perp}}{2}\sum_{i=1}^{N}\left(\gamma_{P_{i},m_{i}^{\prime}}\delta_{\vec{m}^{\prime}+1_{i},\vec{m}}+\gamma_{P_{i},m_{i}}\delta_{\vec{m}+1_{i},\vec{m}^{\prime}}\right)
\end{eqnarray}
Then inserting this result in the calculation of the $c_{M,\vec{m}}^{\left(n+1\right)}\left(t\right)$
we get the next expression for the different orders of the expansion:
\begin{eqnarray}
\dot{c}_{M,\vec{m}}^{\left(n+1\right)}\left(t\right) & = & \frac{i}{2}\sum_{i=1}^{N}\gamma_{P_{i},m_{i}-1}c_{M,\vec{m}-1_{i}}^{\left(n\right)}\left(t\right)e^{-it\left(E_{\vec{m}-1_{i}}^{M}-E_{\vec{m}}^{M}\right)}\nonumber \\
 &  & +\frac{i}{2}\sum_{i=1}^{N}\gamma_{P_{i},m_{i}}c_{M,\vec{m}+1_{i}}^{\left(n\right)}\left(t\right)e^{-it\left(E_{\vec{m}+1_{i}}^{M}-E_{\vec{m}}^{M}\right)}
\end{eqnarray}
To first order in $\Delta_{\perp}$, the solution adds small amplitude
corrections of order $\Delta_{\perp}$ to the unperturbed solution,
by coupling the initial state to all bath configurations where one
bath spin has changed by a unit. To second order in $\Delta_{\perp}$
the solution is more involved, as it contains small amplitude corrections
of order $\Delta_{\perp}^{2}$, plus a secular term:
\begin{equation}
i\left(t-t_{0}\right)c_{M,\vec{m}}^{\left(0\right)}\left(t_{0}\right)\left(\frac{\Delta_{\perp}}{2}\right)^{2}\sum_{i=1}^{N}\left(\frac{\gamma_{P_{i},m_{i}-1}^{2}}{E_{\vec{m}}^{M}-E_{\vec{m}-1_{i}}^{M}}+\frac{\gamma_{P_{i},m_{i}}^{2}}{E_{\vec{m}}^{M}-E_{\vec{m}+1_{i}}^{M}}\right)
\end{equation}
As it is discussed in the main text, secular terms can be renormalized
and give rise to non-perturbative corrections. The main idea is to
assume that the boundary conditions are time-dependent, and that their
series expansion produces the secular terms previously found. This
leads to the next flow equation for the boundary condition:
\begin{equation}
\partial_{\tau}c_{M,\vec{m}}^{\left(0\right)}\left(\tau\right)\simeq ic_{M,\vec{m}}^{\left(0\right)}\left(\tau\right)\left(\frac{\Delta_{\perp}}{2}\right)^{2}\sum_{i=1}^{N}\left(\frac{\gamma_{P_{i},m_{i}-1}^{2}}{E_{\vec{m}}^{M}-E_{\vec{m}-1_{i}}^{M}}+\frac{\gamma_{P_{i},m_{i}}^{2}}{E_{\vec{m}}^{M}-E_{\vec{m}+1_{i}}^{M}}\right)
\end{equation}
with solution
\begin{equation}
c_{M,\vec{m}}^{\left(0\right)}\left(t\right)=c_{M,\vec{m}}^{\left(0\right)}\left(t_{0}\right)\prod_{i=1}^{N}e^{i\left(t-t_{0}\right)\left(\frac{\Delta_{\perp}}{2}\right)^{2}\left(\frac{\gamma_{P_{i},m_{i}-1}^{2}}{E_{\vec{m}-1_{i}}^{M}-E_{\vec{m}}^{M}}+\frac{\gamma_{P_{i},m_{i}}^{2}}{E_{\vec{m}+1_{i}}^{M}-E_{\vec{m}}^{M}}\right)}
\end{equation}
This solution implies that the renormalized solution will oscillate
with a shifted frequency due to the dynamical bath, but it does not
affect the amplitude. Importantly, one must notice that the shift
in frequency could not be obtained perturbatively; however, the instanton
transition is not captured.

\section{Dynamical RG analysis of Mean Field equations}

The general equations of motion for the system are given by:
\begin{eqnarray}
\partial_{t}S^{\alpha} & = & \sum_{\mu}\epsilon_{\mu\alpha\theta}\left(B_{\mu}-\sum_{i=1}^{N}A_{i}^{\mu}I_{i}^{\mu}\right)S^{\theta}\\
\partial_{t}I_{i}^{\alpha} & = & \sum_{\mu}\epsilon_{\mu\alpha\theta}\left(\Delta_{\mu}-S^{\mu}A_{i}^{\mu}\right)I_{i}^{\theta}
\end{eqnarray}
Making use of the mean field decoupling for the statistical averages,
they reduce to:
\begin{eqnarray}
\partial_{t}\langle S^{\alpha}\rangle & = & \sum_{\mu}\epsilon_{\mu\alpha\theta}\left(B_{\mu}-\sum_{i=1}^{N}A_{i}^{\mu}\langle I_{i}^{\mu}\rangle\right)\langle S^{\theta}\rangle\\
\partial_{t}\langle I_{i}^{\alpha}\rangle & = & \epsilon\sum_{\mu}\epsilon_{\mu\alpha\theta}\left(\Delta_{\mu}-\langle S^{\mu}\rangle A_{i}^{\mu}\right)\langle I_{i}^{\theta}\rangle
\end{eqnarray}
where we have introduced the parameter $\epsilon$ to organize the
different powers of perturbations (do not confuse with the Levi-Civita
symbol $\epsilon_{\mu\alpha\theta}$). To lowest order in $\epsilon$,
the equations of motion yield:
\begin{eqnarray}
\partial_{t}\langle S^{\alpha}\rangle_{0} & = & \sum_{\mu}\epsilon_{\mu\alpha\theta}\left(B_{\mu}-\sum_{i=1}^{N}A_{i}^{\mu}m_{i}^{\mu}\right)\langle S^{\theta}\rangle_{0}\\
\partial_{t}\langle I_{i}^{\alpha}\rangle_{0} & = & 0\rightarrow\langle I_{i}^{\alpha}\left(t_{0}\right)\rangle_{0}=m_{i}^{\alpha}
\end{eqnarray}
where $m_{i}^{\alpha}$ are the initial conditions for the magnetization
of each bath spin. The solutions are easily obtained by direct integration,
and the ones for the central spin can be used to calculate the first
order corrections to the bath spin dynamics:
\begin{equation}
\partial_{t}\langle I_{i}^{\alpha}\rangle_{1}=\epsilon\sum_{\mu}\epsilon_{\mu\alpha\theta}\left(\Delta_{\mu}-\langle S^{\mu}\rangle_{0}A_{i}^{\mu}\right)m_{i}^{\theta}
\end{equation}
They solutions display fast oscillations with frequency $\Omega_{\vec{m}}=\sqrt{\sum_{\mu}\left(B_{\mu}-\sum_{i=1}^{N}A_{i}^{\mu}m_{i}^{\mu}\right)^{2}}$,
and the next secular terms:
\begin{eqnarray}
\langle I_{i}^{\alpha}\rangle_{1} & = & \ldots-t\sum_{\nu,\rho}\epsilon_{\alpha\nu\rho}m_{i}^{\nu}\left(\frac{A_{i}^{\rho}\omega_{\rho}}{\Omega_{\vec{m}}^{2}}\sum_{\mu}M_{\mu}\omega_{\mu}-\Delta_{\rho}\right)\label{eq:SecBath}
\end{eqnarray}
 where $\omega_{\mu}=B_{\mu}-\sum_{i=1}^{N}A_{i}^{\mu}m_{i}^{\mu}$
and $M_{\mu}$ is the initial condition for the central spin magnetization
along the $\mu$-axis. The lowest order solution for the central spin
describes a spin precessing in an effective magnetic field, combination
of the external one $\vec{B}$ and the Overhauser field produced by
the static bath $\sum_{i}A_{i}^{\mu}m_{i}^{\mu}$. In order to eliminate
the secular terms for the bath equations of motion to first order
(Eq.\ref{eq:SecBath}), one must consider the boundary conditions
time-dependent, in such a way that they become the generators of the
secular terms to first order. This leads to:
\begin{equation}
\partial_{\tau}m_{i}^{\alpha}\left(\tau\right)=-\sum_{\nu,\rho}\epsilon_{\alpha\nu\rho}m_{i}^{\nu}\left(\tau\right)\left(\frac{A_{i}^{\rho}\omega_{\rho}\left(\tau\right)}{\Omega_{\vec{m}}\left(\tau\right)^{2}}\sum_{\mu}M_{\mu}\left(\tau\right)\omega_{\mu}\left(\tau\right)-\Delta_{\rho}\right)\label{eq:bath-flow1}
\end{equation}
where the $\tau$ dependence due to the boundary conditions has been
explicitly added for clarity. Notice that this is a highly non-linear
differential equation. As the flow equations for the boundary conditions
$m_{i}^{\mu}\left(\tau\right)$ are coupled to the boundary conditions
for the central spin $M_{\mu}\left(\tau\right)$, one must solve the
equations of motion for the central spin to first order in $\epsilon$.
The equations of motion for the central spin, to first order in $\epsilon$,
are:
\begin{eqnarray}
\partial_{t}\langle S^{\alpha}\rangle_{1} & = & \sum_{\mu}\epsilon_{\mu\alpha\theta}\omega_{\mu}\langle S^{\theta}\rangle_{1}-\sum_{\mu}\epsilon_{\mu\alpha\theta}\sum_{i=1}^{N}A_{i}^{\mu}\langle I_{i}^{\mu}\rangle_{1}\langle S^{\theta}\rangle_{0}
\end{eqnarray}
At this point it is useful to define $\eta_{\mu}=\sum_{i}A_{i}^{z}m_{i}^{\mu}=\sum_{i}A_{i}m_{i}^{\mu}$.
The analytical solutions display secular terms, and for clarity, we
derive in detail one of the flow equations for the central spin. For
example, the solution for the $\langle S^{x}\rangle$ component is
given by:
\begin{eqnarray}
\langle S^{x}\left(t\right)\rangle_{1} & = & \epsilon B_{\perp}\Delta_{\perp}\eta_{y}\frac{B_{\perp}M_{y}\left[\cos\left(\Omega_{\vec{m}}t\right)-1\right]+M_{z}\Omega_{\vec{m}}\sin\left(\Omega_{\vec{m}}t\right)}{\Omega_{\vec{m}}^{4}}\label{eq:SecularSx}\\
 &  & -\epsilon tB_{\perp}\Delta_{\perp}\eta_{y}\frac{\omega_{z}\left(B_{\perp}M_{x}+M_{z}\omega_{z}\right)+B_{\perp}\left(B_{\perp}M_{z}-M_{x}\omega_{z}\right)\cos\left(\Omega_{\vec{m}}t\right)-B_{\perp}M_{y}\Omega_{\vec{m}}\sin\left(\Omega_{\vec{m}}t\right)}{\Omega_{\vec{m}}^{4}}\nonumber \\
 &  & +\epsilon t^{2}\Delta_{\perp}\eta_{y}\omega_{z}^{2}\frac{M_{y}\Omega_{\vec{m}}\cos\left(\Omega_{\vec{m}}t\right)+\left(B_{\perp}M_{z}-M_{x}\omega_{z}\right)\sin\left(\Omega_{\vec{m}}t\right)}{2\Omega_{\vec{m}}^{3}}\nonumber 
\end{eqnarray}
Noticing that the unperturbed solution is given by:
\begin{equation}
\langle S^{x}\left(t\right)\rangle_{0}=\frac{B_{\perp}\left(B_{\perp}M_{x}+M_{z}\omega_{z}\right)-\omega_{z}\left(B_{\perp}M_{z}-M_{x}\omega_{z}\right)\cos\left(\Omega_{\vec{m}}t\right)+M_{y}\Omega_{\vec{m}}\omega_{z}\sin\left(\Omega_{\vec{m}}t\right)}{\Omega_{\vec{m}}^{2}}
\end{equation}
one can derive three different flow equations: One for the cosine
term, another for the sine term and a third one corresponding to the
constant term. Let us derive the constant term flow equation, although
the other ones are derived in exactly the same manner. Our aim is
that the boundary condition for the constant term in $\langle S^{x}\left(t\right)\rangle_{0}$
encodes, to first order in $\epsilon$, the secular term obtained
in $\langle S^{x}\left(t\right)\rangle_{1}$. This imposes that:
\begin{equation}
B_{\perp}\partial_{\tau}\frac{B_{\perp}M_{x}\left(\tau\right)+M_{z}\left(\tau\right)\omega_{z}\left(\tau\right)}{\Omega_{\vec{m}}\left(\tau\right)^{2}}=-\epsilon B_{\perp}\Delta_{\perp}\eta_{y}\left(\tau\right)\omega_{z}\left(\tau\right)\frac{B_{\perp}M_{x}\left(\tau\right)+M_{z}\left(\tau\right)\omega_{z}\left(\tau\right)}{\Omega_{\vec{m}}\left(\tau\right)^{4}}\label{eq:Flow1}
\end{equation}
Notice that the secular terms with harmonic time dependence cannot
enter this flow equation, as it must be fulfilled for arbitrary time.
Now one just needs to derive the l.h.s. of Eq.\ref{eq:Flow1} using:
\begin{eqnarray}
\partial_{\tau}\frac{1}{\Omega_{\vec{m}}\left(\tau\right)^{2}} & = & -\frac{2\omega_{z}\left(\tau\right)}{\Omega_{\vec{m}}\left(\tau\right)^{4}}\partial_{\tau}\omega_{z}\left(\tau\right)\nonumber \\
 & = & -2\epsilon\Delta_{\perp}\frac{\omega_{z}\left(\tau\right)\eta_{y}\left(\tau\right)}{\Omega_{\vec{m}}\left(\tau\right)^{4}}
\end{eqnarray}
where we have applied:
\begin{eqnarray}
\partial_{\tau}\omega_{z}\left(\tau\right) & = & -\sum_{i}A_{i}\partial_{\tau}m_{i}^{z}\left(\tau\right)\\
 & = & \epsilon\Delta_{\perp}\sum_{i}A_{i}m_{i}^{y}\left(\tau\right)=\epsilon\Delta_{\perp}\eta_{y}\left(\tau\right)\nonumber 
\end{eqnarray}
and in the last line also Eq.\ref{eq:bath-flow1} for $\partial_{\tau}m_{i}^{z}\left(\tau\right)$.
The flow equation finally becomes:
\begin{equation}
\partial_{\tau}\left(B_{\perp}M_{x}+M_{z}\omega_{z}\right)=\epsilon\Delta_{\perp}\omega_{z}\eta_{y}\frac{B_{\perp}M_{x}+M_{z}\omega_{z}}{\Omega_{\vec{m}}^{2}}
\end{equation}
Finally, reorganizing terms one can write:
\begin{equation}
\partial_{\tau}\log\left(B_{\perp}M_{x}+M_{z}\omega_{z}\right)=\epsilon\Delta_{\perp}\frac{\eta_{y}\omega_{z}}{\Omega_{\vec{m}}^{2}}
\end{equation}
which indicates that the running of the initial condition is triggered
by the bath dynamics, as it is proportional to $\Delta_{\perp}$.
The other equations are derived in a similar fashion, with the peculiarity
that quadratic secular terms $\propto\epsilon t^{2}$ in Eq.\ref{eq:SecularSx}
are also present for the sine and cosine terms. Their appearance is
easy to understand, as they are a consequence of the time-dependent
frequency $\Omega_{\vec{m}}\left(\tau\right)$. Thus, calculating
their flow equation one finds:
\begin{eqnarray}
\partial_{\tau}\log\left(B_{\perp}M_{z}-M_{x}\omega_{z}\right) & = & \epsilon\Delta_{\perp}\frac{\eta_{y}\omega_{z}}{\Omega_{\vec{m}}^{2}}
\end{eqnarray}
and
\begin{eqnarray}
\partial_{\tau}M_{y} & = & 0
\end{eqnarray}
The numerical solution of the flow equations captures the instanton-like
dynamics, with the resonant transition happening when the bath depolarizes
(Fig.\ref{fig:Fig1}). The comparison with the exact diagonalization
result shows that the moment where the instanton transitions happen
is well captured, however a slow decay takes over at very long time-scales.
On the other hand, the suppression of coherent oscillations at short
times is missed by this solution. This will be captured by adding
correlations in the next section.

\begin{figure}
\includegraphics[scale=0.45]{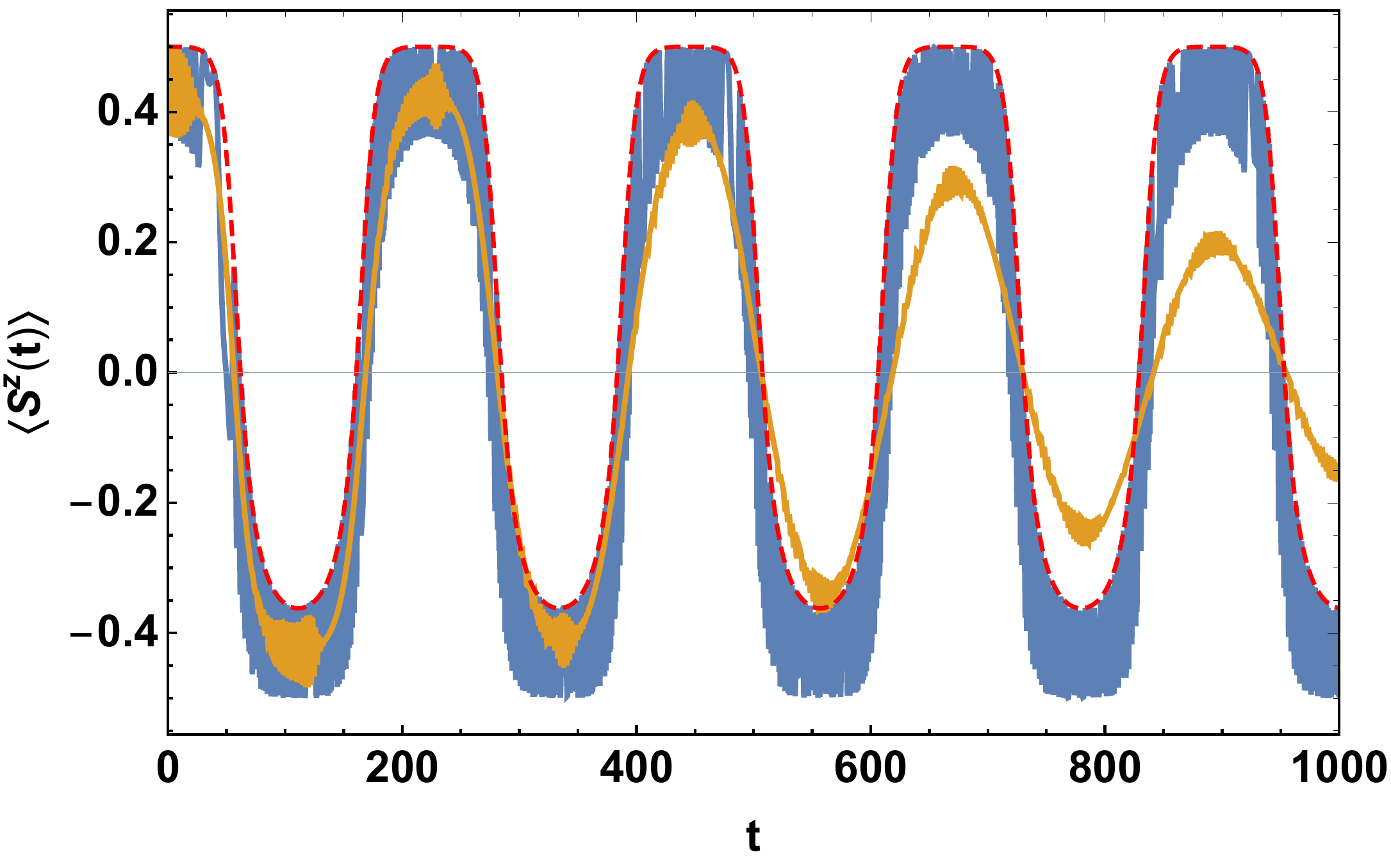}

\caption{\label{fig:Fig1}Comparison between the central spin dynamics between
the dRG result to first order (blue) and the exact diagonalization
one (yellow). The red dashed line shows the function $M_{z}\left(t\right)$,
which modulates the fast coherent oscillations of the central spin.
Parameters: $\Delta_{\perp}/B_{\perp}=0.03$, $A/B_{\perp}=0.05$
and $N=100$ for the initial condition $\langle S^{z}\left(0\right)\rangle=\frac{1}{2}$
and $\langle P_{z}\left(0\right)\rangle=N/2$.}
\end{figure}

\section{Correlations: comparison for different decoupling methods}

When correlations between the system and the bath spins are included,
one needs to calculate the next equation of motion:
\begin{eqnarray}
\partial_{t}I_{n}^{\beta}S^{\alpha} & = & i\left[H,I_{n}^{\beta}S^{\alpha}\right]\\
 & = & \epsilon_{z\alpha\mu}B_{z}I_{n}^{\beta}S^{\mu}+\epsilon_{x\alpha\mu}B_{\perp}I_{n}^{\beta}S^{\mu}\nonumber \\
 &  & +\epsilon_{z\beta\nu}\Delta_{z}I_{n}^{\nu}S^{\alpha}+\epsilon_{x\beta\nu}\Delta_{\perp}I_{n}^{\nu}S^{\alpha}\nonumber \\
 &  & -\epsilon_{z\beta\nu}A_{n}I_{n}^{\nu}S^{z}S^{\alpha}-\epsilon_{z\alpha\mu}\sum_{i=1}^{N}A_{i}I_{n}^{\beta}I_{i}^{z}S^{\mu}\nonumber 
\end{eqnarray}
It is important to now separate the terms $i=n$ and $i\neq n$ due
their different effects. This leads to the next equation of motion:
\begin{eqnarray}
\partial_{t}I_{n}^{\beta}S^{\alpha} & = & \left[\epsilon_{x\alpha\mu}B_{\perp}+\epsilon_{z\alpha\mu}\left(B_{z}-\sum_{i\neq n}^{N}A_{i}I_{i}^{z}\right)\right]I_{n}^{\beta}S^{\mu}\nonumber \\
 &  & +\left(\epsilon_{z\beta\nu}\Delta_{z}+\epsilon_{x\beta\nu}\Delta_{\perp}\right)I_{n}^{\nu}S^{\alpha}\nonumber \\
 &  & -\epsilon_{z\beta\nu}A_{n}I_{n}^{\nu}\left(\frac{\delta_{z,\alpha}}{4}+i\epsilon_{z\alpha\mu}S^{\mu}\right)-\epsilon_{z\alpha\mu}A_{n}I_{n}^{\beta}I_{n}^{z}S^{\mu}
\end{eqnarray}
Furthermore, if one considers that the bath spins are spin $1/2$,
the expression simplifies to:
\begin{eqnarray}
\partial_{t}I_{n}^{\beta}S^{\alpha} & = & \left[\epsilon_{x\alpha\mu}B_{\perp}+\epsilon_{z\alpha\mu}\left(B_{z}-\sum_{i\neq n}^{N}A_{i}I_{i}^{z}\right)\right]I_{n}^{\beta}S^{\mu}\label{eq:Correlator1}\\
 &  & +\left(\epsilon_{z\beta\nu}\Delta_{z}+\epsilon_{x\beta\nu}\Delta_{\perp}\right)I_{n}^{\nu}S^{\alpha}-\frac{A_{n}}{4}\left(\epsilon_{z\beta\nu}\delta_{z,\alpha}I_{n}^{\nu}+\epsilon_{z\alpha\mu}\delta_{z,\beta}S^{\mu}\right)\nonumber 
\end{eqnarray}
We now describe three different decoupling schemes that have been
used to study their accuracy for the simulation of dynamics (Fig.\ref{fig:Comparison-decoupling}).
\begin{figure}
\includegraphics[scale=0.6]{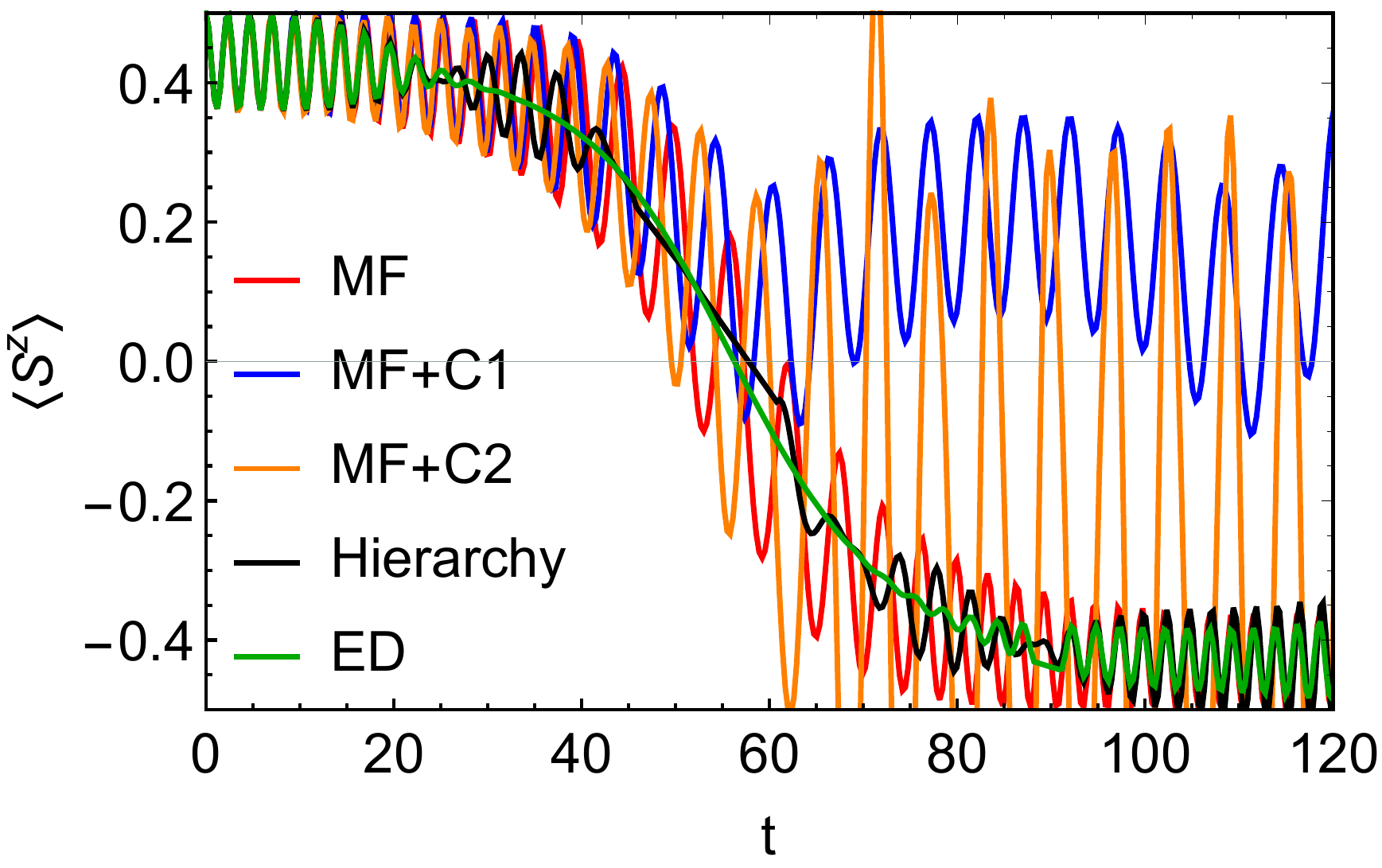}

\caption{\label{fig:Comparison-decoupling}Comparison between different decoupling
schemes}
\end{figure}
 The first one consists in neglecting all correlations in Eq.\ref{eq:Correlator1},
which leads to:
\begin{eqnarray}
\partial_{t}\langle I_{n}^{\beta}S^{\alpha}\rangle & \simeq & \left[\epsilon_{x\alpha\mu}B_{\perp}+\epsilon_{z\alpha\mu}\left(B_{z}-\sum_{i\neq n}^{N}A_{i}\langle I_{i}^{z}\rangle\right)\right]\langle I_{n}^{\beta}\rangle\langle S^{\mu}\rangle\nonumber \\
 &  & +\left(\epsilon_{z\beta\nu}\Delta_{z}+\epsilon_{x\beta\nu}\Delta_{\perp}\right)\langle I_{n}^{\nu}\rangle\langle S^{\alpha}\rangle\nonumber \\
 &  & -\frac{A_{n}}{4}\left(\epsilon_{z\beta\nu}\delta_{z,\alpha}\langle I_{n}^{\nu}\rangle+\epsilon_{z\alpha\mu}\delta_{z,\beta}\langle S^{\mu}\rangle\right)
\end{eqnarray}
Solving these equations, simultaneously with the ones for the central
spin, results in the solutions labeled as MF+C1 in Fig.\ref{fig:Comparison-decoupling}.
Unfortunately they are in disagreement with the exact result and produce
an even worse approximation than the MF solution: The instanton transitions
are absent, and the suppression of coherent oscillations not captured.
The reason is that we have kept certain correlated parts, but eliminated
others without a physical criteria, introducing unphysical non-linear
terms. It is important to understand that small changes in the non-linear
terms, due to how the equations are truncated, will produce small
differences at short time, but in general they can be dominant at
long time. This is one of the reasons why simulating dynamics can
be quite complicated.

If we consider a more complex decoupling, where two-spin correlations
are maintained, we can just separate the three-spin correlators as
$\langle I_{i}^{z}I_{n}^{\beta}S^{\mu}\rangle\simeq\langle I_{i}^{z}\rangle\langle I_{n}^{\beta}S^{\mu}\rangle$,
which mostly assumes that when fluctuations are small around $\langle I_{i}^{z}\rangle$,
the solution should be accurate. The next equation is then obtained
for the two-spin function:
\begin{eqnarray}
\partial_{t}\langle I_{n}^{\beta}S^{\alpha}\rangle & \simeq & \left[\epsilon_{x\alpha\mu}B_{\perp}+\epsilon_{z\alpha\mu}\left(B_{z}-\sum_{i\neq n}^{N}A_{i}\langle I_{i}^{z}\rangle\right)\right]\langle I_{n}^{\beta}S^{\mu}\rangle\nonumber \\
 &  & +\left(\epsilon_{z\beta\nu}\Delta_{z}+\epsilon_{x\beta\nu}\Delta_{\perp}\right)\langle I_{n}^{\nu}S^{\alpha}\rangle\nonumber \\
 &  & -\frac{A_{n}}{4}\left(\epsilon_{z\beta\nu}\delta_{z,\alpha}\langle I_{n}^{\nu}\rangle+\epsilon_{z\alpha\mu}\delta_{z,\beta}\langle S^{\mu}\rangle\right)
\end{eqnarray}
This gives the numerical results labeled as MF+C2 in the main text.
The number of coupled equations is now larger, and there is back-reaction
included between the different two-point correlators. Nevertheless
the numerical solution is still at large disagreement with the exact
solution.

Finally, an alternative decoupling is obtained by separating the n-point
functions into correlated and uncorrelated parts. Here, one must determine
the equation of motion for the correlated part $\partial_{t}\langle I_{n}^{\beta}S^{\alpha}\rangle^{c}=\partial_{t}\langle I_{n}^{\beta}S^{\alpha}\rangle-\partial_{t}\langle I_{n}^{\beta}\rangle\langle S^{\alpha}\rangle$:
\begin{eqnarray}
\partial_{t}\langle I_{n}^{\beta}S^{\alpha}\rangle^{c} & = & \epsilon_{z\alpha\mu}B_{z}\langle I_{n}^{\beta}S^{\mu}\rangle^{c}+\epsilon_{x\alpha\mu}B_{\perp}\langle I_{n}^{\beta}S^{\mu}\rangle^{c}\\
 &  & +\epsilon_{z\beta\nu}\Delta_{z}\langle I_{n}^{\nu}S^{\alpha}\rangle^{c}+\epsilon_{x\beta\nu}\Delta_{\perp}\langle I_{n}^{\nu}S^{\alpha}\rangle^{c}\nonumber \\
 &  & -\epsilon_{z\alpha\mu}\sum_{i\neq n}^{N}A_{i}\left(\langle S^{\mu}\rangle\langle I_{n}^{\beta}I_{i}^{z}\rangle^{c}+\langle I_{i}^{z}\rangle\langle I_{n}^{\beta}S^{\mu}\rangle^{c}+\langle I_{n}^{\beta}I_{i}^{z}S^{\mu}\rangle^{c}\right)\nonumber \\
 &  & -\epsilon_{z\beta\nu}A_{n}\langle I_{n}^{\nu}\rangle\left(\frac{\delta_{z,\alpha}}{4}-\langle S^{\alpha}\rangle\langle S^{z}\rangle\right)\nonumber \\
 &  & -\epsilon_{z\alpha\mu}A_{n}\left(\frac{\delta_{z,\beta}}{4}-\langle I_{n}^{\beta}\rangle\langle I_{n}^{z}\rangle\right)\langle S^{\mu}\rangle\nonumber \\
 &  & +A_{n}\left(\epsilon_{z\beta\nu}\langle S^{\alpha}\rangle\langle I_{n}^{\nu}S^{z}\rangle^{c}+\epsilon_{z\alpha\mu}\langle I_{n}^{\beta}\rangle\langle I_{n}^{z}S^{\mu}\rangle^{c}\right)\nonumber 
\end{eqnarray}
The previous equation is exact but couples to higher spin correlators.
Now we assume that three-point correlation functions are subdominant,
and they can be neglected (at least at short time). This results in:
\begin{eqnarray}
\partial_{t}\langle I_{n}^{\beta}S^{\alpha}\rangle^{c} & \simeq & \epsilon_{z\alpha\mu}B_{z}\langle I_{n}^{\beta}S^{\mu}\rangle^{c}+\epsilon_{x\alpha\mu}B_{\perp}\langle I_{n}^{\beta}S^{\mu}\rangle^{c}\label{eq:Corr0}\\
 &  & +\epsilon_{z\beta\nu}\Delta_{z}\langle I_{n}^{\nu}S^{\alpha}\rangle^{c}+\epsilon_{x\beta\nu}\Delta_{\perp}\langle I_{n}^{\nu}S^{\alpha}\rangle^{c}\nonumber \\
 &  & -\epsilon_{z\alpha\mu}\sum_{i\neq n}^{N}A_{i}\left(\langle S^{\mu}\rangle\langle I_{n}^{\beta}I_{i}^{z}\rangle^{c}+\langle I_{i}^{z}\rangle\langle I_{n}^{\beta}S^{\mu}\rangle^{c}\right)\nonumber \\
 &  & -\epsilon_{z\beta\nu}A_{n}\langle I_{n}^{\nu}\rangle\left(\frac{\delta_{z,\alpha}}{4}-\langle S^{\alpha}\rangle\langle S^{z}\rangle\right)\nonumber \\
 &  & -\epsilon_{z\alpha\mu}A_{n}\left(\frac{\delta_{z,\beta}}{4}-\langle I_{n}^{\beta}\rangle\langle I_{n}^{z}\rangle\right)\langle S^{\mu}\rangle\nonumber \\
 &  & +A_{n}\left(\epsilon_{z\beta\nu}\langle S^{\alpha}\rangle\langle I_{n}^{\nu}S^{z}\rangle^{c}+\epsilon_{z\alpha\mu}\langle I_{n}^{\beta}\rangle\langle I_{n}^{z}S^{\mu}\rangle^{c}\right)\nonumber 
\end{eqnarray}
These set of equations can be solved numerically and they give quite
accurate results. However, to capture the suppression of the oscillations
observed from exact diagonalization, it is enough to consider a simpler
approximation of the previous equation, by neglecting the slow terms
given by the bath-bath correlators $\langle I_{n}^{\beta}I_{i}^{z}\rangle^{c}\simeq0$,
the terms proportional to $\vec{\Delta}$, and the two terms in the
last line. This approximation will fail at long times, but reduces
the number of equations considerably. This gives the numerical results
labeled as Hierarchy in the main text (Fig.\ref{fig:Comparison-corr1})
and in Fig.\ref{fig:Comparison-decoupling}.

One important aspect introduced by the correlated parts is the appearance
of quadratic terms $\langle I_{n}^{z}\rangle^{2}$ in the equation
of motion, which even for large disorder, where the bath would be
unpolarized on average, are non-vanishing.

It is important to notice that the correlators entering the equation
of motion for the central spin magnetization are summed over all bath
spins (i.e., they are proportional to $\sum_{i}A_{i}\langle I_{i}^{z}S^{\theta}\rangle^{c}$.
This means that when numerically solving the equations for the correlated
parts, one must solve for $\sum_{n}A_{n}\langle I_{n}^{\beta}S^{\alpha}\rangle^{c}$
and not $\langle I_{n}^{\beta}S^{\alpha}\rangle^{c}$. Therefore,
the quadratic terms $\langle I_{n}^{\beta}\rangle\langle I_{n}^{z}\rangle$
require the addition of their equation of motion, as they characterize
the variance of the bath, which is different to the square of its
average value. However, its derivation is simple, as it can be directly
linked with the previous equations of motion. For example, neglecting
all correlations, they yield:
\begin{eqnarray}
\partial_{t}\langle I_{n}^{\alpha}\rangle\langle I_{n}^{\beta}\rangle & \simeq & \Delta_{\perp}\left(\epsilon_{x\beta\theta}\langle I_{n}^{\alpha}\rangle+\epsilon_{x\alpha\theta}\langle I_{n}^{\beta}\rangle\right)\langle I_{n}^{\theta}\rangle\\
 &  & -A_{n}\langle S^{z}\rangle\left(\epsilon_{z\beta\theta}\langle I_{n}^{\alpha}\rangle+\epsilon_{z\alpha\theta}\langle I_{n}^{\beta}\rangle\right)\langle I_{n}^{\theta}\rangle\nonumber 
\end{eqnarray}
where in the last line we have factorized the system-bath operators.
This is how the equations have been numerically solved in this work.

\section{Flow equations for the Hierarchy of correlations}

To obtain the flow equations that encode the effect of amplitude renormalization
using dRG we consider Eq.\ref{eq:Corr0} with the assumption $\left|\vec{\Delta}\right|,A_{i}\ll\left|\vec{B}\right|$,
as previously discussed. Then the equation has the next leading terms:
\begin{eqnarray}
\partial_{t}\langle I_{n}^{\beta}S^{\alpha}\rangle^{c} & \simeq & \epsilon_{x\alpha\mu}B_{\perp}\langle I_{n}^{\beta}S^{\mu}\rangle^{c}+\epsilon_{z\alpha\mu}\left(B_{z}-\sum_{i\neq n}^{N}A_{i}\langle I_{i}^{z}\rangle\right)\langle I_{n}^{\beta}S^{\mu}\rangle^{c}\label{eq:Corr-dRG-eq}\\
 &  & -\epsilon_{z\alpha\mu}A_{n}\langle S^{\mu}\rangle\left(\frac{\delta_{z,\beta}}{4}-\langle I_{n}^{\beta}\rangle\langle I_{n}^{z}\rangle\right)-\epsilon_{z\beta\nu}A_{n}\langle I_{n}^{\nu}\rangle\left(\frac{\delta_{z,\alpha}}{4}-\langle S^{\alpha}\rangle\langle S^{z}\rangle\right)\nonumber 
\end{eqnarray}
where we have neglected bath-bath correlators (which are proportional
to $\left|\vec{\Delta}\right|$ and therefore sub-dominant) as well
as the precession of the spins in their internal field $\vec{\Delta}$.
The first line in Eq.\ref{eq:Corr-dRG-eq} corresponds to the fastest
time-scale, coming from the central spin dynamics precession, while
the second line is proportional to $A_{n}$ and suppresses the amplitude
of the oscillations as the spins precess away from the longitudinal
axis. For the three relevant components one has:
\begin{eqnarray}
\partial_{t}\langle I_{n}^{z}S^{x}\rangle^{c} & = & \left(B_{z}-\sum_{i\neq n}^{N}A_{i}\langle I_{i}^{z}\rangle\right)\langle I_{n}^{z}S^{y}\rangle^{c}-A_{n}\langle S^{y}\rangle\left(\frac{1}{4}-\langle I_{n}^{z}\rangle^{2}\right)\label{eq:Hierarchy1}\\
\partial_{t}\langle I_{n}^{z}S^{y}\rangle^{c} & = & B_{\perp}\langle I_{n}^{z}S^{z}\rangle^{c}-\left(B_{z}-\sum_{i\neq n}^{N}A_{i}\langle I_{i}^{z}\rangle\right)\langle I_{n}^{z}S^{x}\rangle^{c}+A_{n}\langle S^{x}\rangle\left(\frac{1}{4}-\langle I_{n}^{z}\rangle^{2}\right)\\
\partial_{t}\langle I_{n}^{z}S^{z}\rangle^{c} & = & -B_{\perp}\langle I_{n}^{z}S^{y}\rangle^{c}
\end{eqnarray}
Notice that if we insert the dimensionless parameter $\epsilon$,
to keep track of the slow bath dynamics, the previous equations are
equivalent to an expansion up to linear order in $\epsilon$. At short
times the first line dominates they reduce to:
\begin{eqnarray}
\partial_{t}\langle I_{n}^{z}S^{x}\rangle_{0}^{c} & = & \omega_{z}\langle I_{n}^{z}S^{y}\rangle_{0}^{c}\\
\partial_{t}\langle I_{n}^{z}S^{y}\rangle_{0}^{c} & = & B_{\perp}\langle I_{n}^{z}S^{z}\rangle_{0}^{c}-\omega_{z}\langle I_{n}^{z}S^{x}\rangle_{0}^{c}\\
\partial_{t}\langle I_{n}^{z}S^{z}\rangle_{0}^{c} & = & -B_{\perp}\langle I_{n}^{z}S^{y}\rangle_{0}^{c}
\end{eqnarray}
The solution is similar to the one for the central spin, but with
different boundary conditions, which describe the initial correlations
between system and bath. Furthermore, as one only needs its sum over
all bath spins multiplied by $A_{n}$, we can solve the next equation
of motion instead:
\begin{eqnarray}
\partial_{t}\langle P^{z}S^{x}\rangle_{0}^{c} & = & \omega_{z}\langle P^{z}S^{y}\rangle_{0}^{c}\\
\partial_{t}\langle P^{z}S^{y}\rangle_{0}^{c} & = & B_{\perp}\langle P^{z}S^{z}\rangle_{0}^{c}-\omega_{z}\langle P^{z}S^{x}\rangle_{0}^{c}\\
\partial_{t}\langle P^{z}S^{z}\rangle_{0}^{c} & = & -B_{\perp}\langle P^{z}S^{y}\rangle_{0}^{c}
\end{eqnarray}
where $P^{z}=\sum_{n}A_{n}I_{n}^{z}$, and the solutions are now independent
of the specific bath spin. As we assume that correlated parts are
small corrections to mean field, at least for short time, and we attach
to them an $\epsilon$ factor in the equations of motion for the central
spin and bath spins. Therefore, the lowest order the solutions for
the central spin and the bath are unchanged, as all the differences
due to correlations will happen to linear order in $\epsilon$. To
first order in $\epsilon$ the equations of motion for the bath spins
are:
\begin{eqnarray}
\partial_{t}\langle I_{i}^{x}\rangle_{1} & = & \epsilon\left(\Delta_{z}-A_{i}\langle S^{z}\rangle_{0}\right)m_{i}^{y}\\
\partial_{t}\langle I_{i}^{y}\rangle_{1} & = & \epsilon\Delta_{\perp}m_{i}^{z}-\epsilon\left(\Delta_{z}-A_{i}\langle S^{z}\rangle_{0}\right)m_{i}^{x}\\
\partial_{t}\langle I_{i}^{z}\rangle_{1} & = & -\epsilon\Delta_{\perp}m_{i}^{y}
\end{eqnarray}
where we have neglected the correlated terms because they are of order
$\epsilon^{2}$. For the central spin, to first order in $\epsilon$,
one finds:
\begin{eqnarray}
\partial_{t}\langle S^{x}\rangle_{1} & = & \omega_{z}\langle S^{y}\rangle_{1}-\langle S^{y}\rangle_{0}\sum_{i}A_{i}\langle I_{i}^{z}\rangle_{1}-\epsilon\sum_{i}A_{i}\langle I_{i}^{z}S^{y}\rangle_{0}^{c}\\
\partial_{t}\langle S^{y}\rangle_{1} & = & B_{\perp}\langle S^{z}\rangle_{1}-\omega_{z}\langle S^{x}\rangle_{1}+\langle S^{x}\rangle_{0}\sum_{i}A_{i}\langle I_{i}^{z}\rangle_{1}+\epsilon\sum_{i}A_{i}\langle I_{i}^{z}S^{x}\rangle_{0}^{c}\\
\partial_{t}\langle S^{z}\rangle_{1} & = & -B_{\perp}\langle S^{y}\rangle_{1}
\end{eqnarray}
and for the correlations:
\begin{eqnarray}
\partial_{t}\langle I_{n}^{z}S^{x}\rangle_{1}^{c} & = & \omega_{z}\langle I_{n}^{z}S^{y}\rangle_{1}^{c}-\langle I_{n}^{z}S^{y}\rangle_{0}^{c}\sum_{i\neq n}^{N}A_{i}\langle I_{i}^{z}\rangle_{1}-\epsilon A_{n}\langle S^{y}\rangle_{0}\left(\frac{1}{4}-\left(m_{n}^{z}\right)^{2}\right)\\
\partial_{t}\langle I_{n}^{z}S^{y}\rangle_{1}^{c} & = & B_{\perp}\langle I_{n}^{z}S^{z}\rangle_{1}^{c}-\omega_{z}\langle I_{n}^{z}S^{x}\rangle_{1}^{c}+\langle I_{n}^{z}S^{x}\rangle_{0}^{c}\sum_{i\neq n}^{N}A_{i}\langle I_{i}^{z}\rangle_{1}+\epsilon A_{n}\langle S^{x}\rangle_{0}\left(\frac{1}{4}-\left(m_{n}^{z}\right)^{2}\right)\\
\partial_{t}\langle I_{n}^{z}S^{z}\rangle_{1}^{c} & = & -B_{\perp}\langle I_{n}^{z}S^{y}\rangle_{1}^{c}
\end{eqnarray}
The appearance of new secular terms in the different solutions is
what gives rise to the modified flow equations for the system. For
the bath, the flow equations are identical to the mean-field case,
because we have neglected the role of correlations on it:
\begin{eqnarray}
\partial_{\tau}m_{i}^{x} & = & m_{i}^{y}\left(\Delta_{z}-A_{i}\omega_{z}\frac{B_{\perp}M_{x}+M_{z}\omega_{z}}{\Omega_{\vec{m}}^{2}}\right)\\
\partial_{\tau}m_{i}^{y} & = & m_{i}^{z}\Delta_{\perp}-m_{i}^{x}\left(\Delta_{z}-A_{i}\omega_{z}\frac{B_{\perp}M_{x}+M_{z}\omega_{z}}{\Omega_{\vec{m}}^{2}}\right)\\
\partial_{\tau}m_{i}^{z} & = & -m_{i}^{y}\Delta_{\perp}\label{eq:EOM-mz}
\end{eqnarray}
The solutions for the bath to first order will now be used in the
calculation for the central spin, which requires to multiply by $A_{i}$
and sum over all bath spins $\sum_{i}A_{i}\langle I_{i}^{z}\rangle_{1}=-\epsilon\left(t-t_{0}\right)\Delta_{\perp}\sum_{i}A_{i}m_{i}^{y}=-\epsilon\left(t-t_{0}\right)\Delta_{\perp}\eta_{y}$.
The same needs to be done for the lowest order solutions of the correlated
parts, which is why we have defined $a_{\alpha\beta}=\sum_{i}A_{i}c_{i}^{\alpha\beta}$,
and $c_{i}^{\alpha\beta}$ is defined as the initial condition for
the correlated part $\langle I_{i}^{\alpha}S^{\beta}\rangle^{c}$.
The equations of motion for the central spin, to first order in $\epsilon$,
become:
\begin{eqnarray}
\partial_{t}\langle S^{x}\rangle_{1} & = & \omega_{z}\langle S^{y}\rangle_{1}+\epsilon\left(t-t_{0}\right)\Delta_{\perp}\eta_{y}\langle S^{y}\rangle_{0}-\epsilon\langle P^{z}S^{y}\rangle_{0}^{c}\\
\partial_{t}\langle S^{y}\rangle_{1} & = & B_{\perp}\langle S^{z}\rangle_{1}-\omega_{z}\langle S^{x}\rangle_{1}-\epsilon\left(t-t_{0}\right)\Delta_{\perp}\eta_{y}\langle S^{x}\rangle_{0}+\epsilon\langle P^{z}S^{x}\rangle_{0}^{c}\\
\partial_{t}\langle S^{z}\rangle_{1} & = & -B_{\perp}\langle S^{y}\rangle_{1}
\end{eqnarray}
and their flow equation yields:
\begin{eqnarray}
\partial_{\tau}\log\left(B_{\perp}M_{x}+M_{z}\omega_{z}\right) & = & \epsilon\Delta_{\perp}\omega_{z}\frac{\eta_{y}}{\Omega_{\vec{m}}^{2}}\\
\partial_{\tau}\left(B_{\perp}M_{z}-\omega_{z}M_{x}\right) & = & \epsilon\Delta_{\perp}\omega_{z}\eta_{y}\frac{B_{\perp}M_{z}-\omega_{z}M_{x}}{\Omega_{\vec{m}}^{2}}+\epsilon\omega_{z}a_{zy}\\
\partial_{\tau}M_{y} & = & \epsilon\omega_{z}\frac{a_{zx}\omega_{z}-B_{\perp}a_{zz}}{\Omega_{\vec{m}}^{2}}
\end{eqnarray}
In absence of correlated parts ($a_{\alpha\beta}=0$) one recovers
the mean field flow equations, and correlations modify the longitudinal
and transverse magnetizations by coupling them with $M_{y}$, which
can now flow. Finally, for the boundary conditions of the correlated
parts, we solve the next equation of motion:
\begin{eqnarray}
\partial_{t}\langle P^{z}S^{x}\rangle_{1}^{c} & = & \omega_{z}\langle P^{z}S^{y}\rangle_{1}^{c}+\epsilon\left(t-t_{0}\right)\Delta_{\perp}\eta_{y}\langle P^{z}S^{y}\rangle_{0}^{c}-\epsilon\langle S^{y}\rangle_{0}\xi_{z}\\
\partial_{t}\langle P^{z}S^{y}\rangle_{1}^{c} & = & B_{\perp}\langle P^{z}S^{z}\rangle_{1}^{c}-\omega_{z}\langle P^{z}S^{x}\rangle_{1}^{c}-\epsilon\left(t-t_{0}\right)\Delta_{\perp}\eta_{y}\langle P^{z}S^{x}\rangle_{0}^{c}+\epsilon\langle S^{x}\rangle_{0}\xi_{z}\\
\partial_{t}\langle P^{z}S^{z}\rangle_{1}^{c} & = & -B_{\perp}\langle P^{z}S^{y}\rangle_{1}^{c}
\end{eqnarray}
where we have defined $\xi_{z}=\sum_{n=1}^{N}A_{n}^{2}\left(\frac{1}{4}-\left(m_{n}^{z}\right)^{2}\right)$.
The solution leads to the next flow equations:
\begin{eqnarray}
\partial_{\tau}\log\left(B_{\perp}a_{zx}+a_{zz}\omega_{z}\right) & = & \epsilon\Delta_{\perp}\omega_{z}\frac{\eta_{y}}{\Omega_{\vec{m}}^{2}}\\
\partial_{\tau}\left(B_{\perp}a_{zz}-a_{zx}\omega_{z}\right) & = & \epsilon\Delta_{\perp}\omega_{z}\frac{\eta_{y}}{\Omega_{\vec{m}}^{2}}\left(B_{\perp}a_{zz}-a_{zx}\omega_{z}\right)+\epsilon\omega_{z}M_{y}\xi_{z}\\
\partial_{\tau}a_{zy} & = & -\epsilon\frac{\xi_{z}\omega_{z}}{\Omega^{2}}\left(B_{\perp}M_{z}-\omega_{z}M_{x}\right)
\end{eqnarray}
Then one just needs to define the disorder correlators for the disordered
case, or directly solve for the ordered case. Notice that in the ordered
case $\xi_{z}=A^{2}\frac{N}{4}-A^{2}\sum_{n}\left(m_{n}^{z}\right)^{2}$,
where now $m_{n}^{z}\left(\tau\right)$ needs to be determined. However,
it is easy to find its equation of motion from Eq.\ref{eq:EOM-mz}. 

Fig.\ref{fig:Supp-Fig2} shows a comparison between the exact numerical
simulation and approximation using the Hierarchy of Correlations decoupling
at long times. It is clear that although the slow decay is not captured,
because it corresponds to a higher order correction in the flow equations,
the agreement is much better than the MF solution and than all the
other approximations previously tried.

\begin{figure}
\includegraphics[scale=0.5]{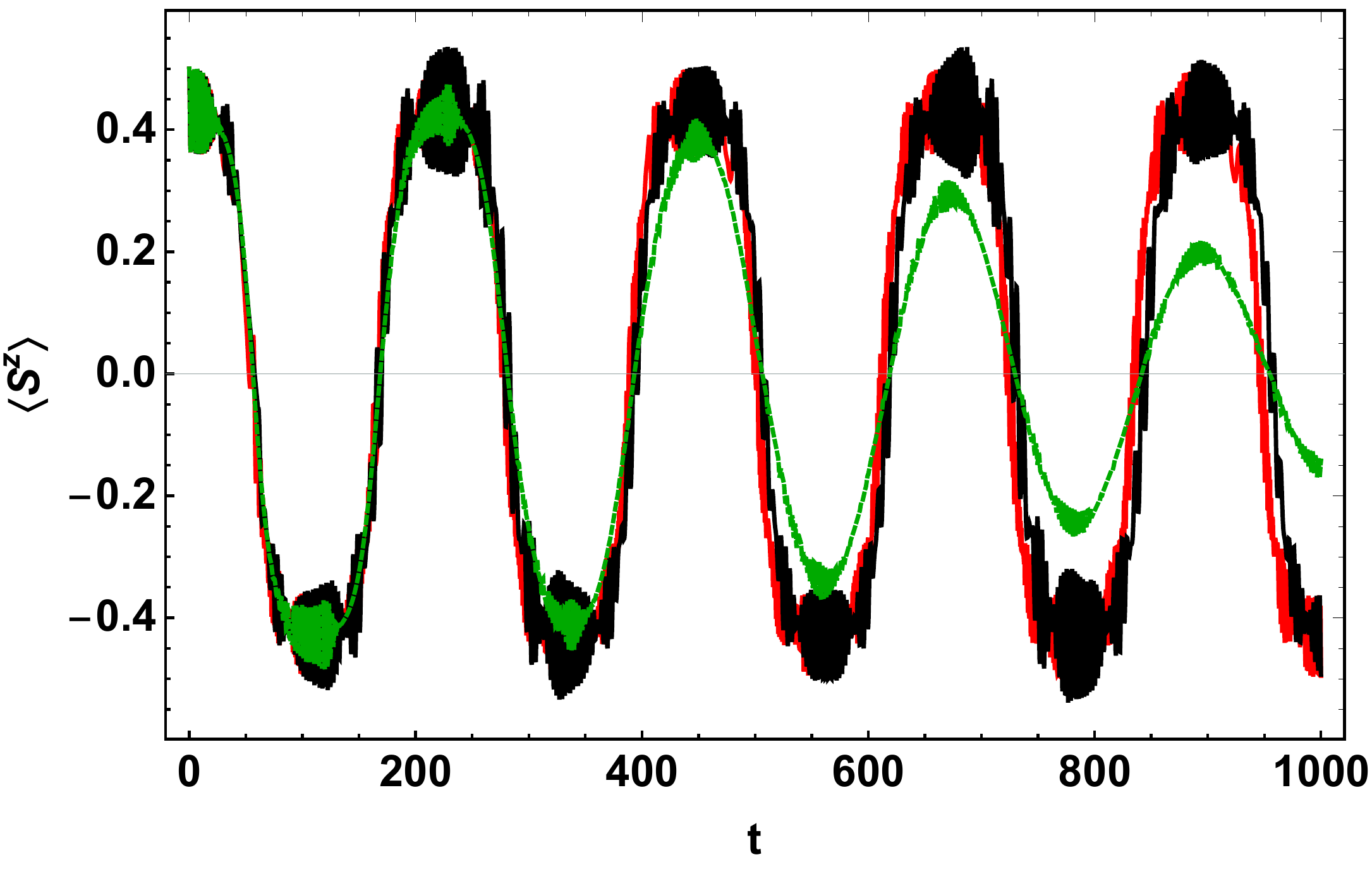}

\caption{\label{fig:Supp-Fig2}Comparison between the exact Hierarchy of correlations
(black), its lowest order approximation using dRG (red) and the exact
dynamics for the central spin (green), at long times. Parameters:
$\Delta/B=0.03$, $J/B=0.05$ and $N=100$ for the initial condition
$\langle S^{z}\left(0\right)\rangle=\frac{1}{2}$ and $\langle P_{z}\left(0\right)\rangle=N/2$.}
\end{figure}
Finally, Fig.\ref{fig:Fig3} shows the solutions from the flow equations
for the cross-correlator between central spin and bath, and the magnetization
components $M_{x,z}\left(t\right)$. It is interesting to see that
the periodicity of the flow equations happens at quite long time (of
the order of $t\sim220$ in units of $B_{\perp}^{-1}$) and that the
cross-correlator initial condition $a_{zy}\left(t\right)$ has large
corrections, even when the system starts in an uncorrelated state
(corrections of the order of $\sim B_{\perp}$ rather than $A_{i}$
or $\Delta_{\perp}$, which is what one would find in absence of renormalization).
This means that central spin and bath become highly correlated and
these correlations are needed for a correct description of the dynamics
at late time.

\begin{figure}
\includegraphics[scale=0.45]{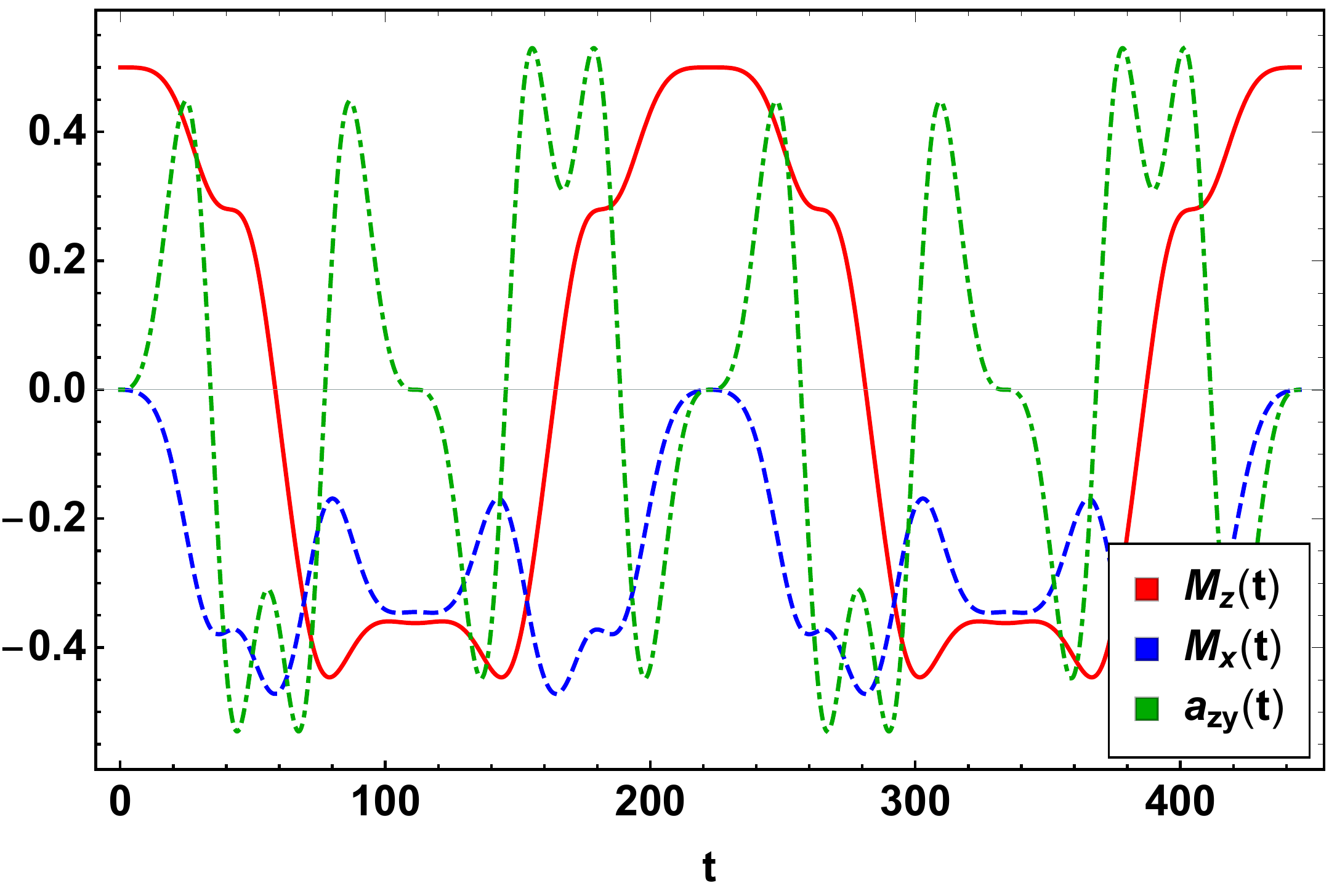}

\caption{\label{fig:Fig3}Numerical solution of the flow equations for the
running of the different initial conditions. (Red, solid line) $M_{z}\left(t\right)$
still displays instanton-like behavior, but the profile is modified
due to correlations with respect to the uncorrelated case (see Fig.\ref{fig:Fig1},
red-dashed line). (Blue, dashed line) $M_{x}\left(t\right)$ also
shows large corrections but it is always negative, which is expected
due to the positive value chosen for the interaction parameter. (Green,
dot-dashed line) $a_{zy}\left(t\right)$ displays complicated oscillations,
correlated with the bath and the central spin magnetization, but importantly,
acquires large values that can largely modify the mean field dynamics.}
\end{figure}
\end{widetext}
\end{document}